\documentstyle[preprint,aps,epsf,tighten]{revtex}  

\newcommand{\slashed}[1]{\rlap{$#1$}/}
\newcommand{\slashp}{\mbox{$\not \hspace*{-1.10mm} p$}}

\newcommand{\GeV}{\mbox{\rm GeV}}
\newcommand{\keV}{\mbox{\rm keV}}
\newcommand{\MeV}{\mbox{\rm MeV}}
\newcommand{\lsim}[1]{
\setlength{\unitlength}{12pt}
\begin{picture}(1.4,1.)
\put(.7,-0.3){\makebox(0.0,1.)[t]{$<$}}
\put(.7,-0.3){\makebox(0.0,1.)[b]{$\sim$}}
\end{picture}#1}

\begin{document}

\draft

\preprint{ZTF-97/R01}


\title{ $\eta$ and $\eta^\prime$ in a coupled Schwinger-Dyson 
              \\ and Bethe-Salpeter approach }

\author{Dubravko Klabu\v{c}ar}
\address{\footnotesize Department of Physics, Faculty of Science, \\
	Zagreb University, P.O.B. 162, 10001 Zagreb, Croatia}
\author{Dalibor Kekez}
\address{\footnotesize Rudjer Bo\v{s}kovi\'{c} Institute,
	 P.O.B. 1016, 10001 Zagreb, Croatia}

\maketitle

\vspace*{-10mm}
\begin{abstract}
\noindent Extending our earlier treatments of 
$\pi^0, \eta_c$ and $\eta_b$, we study the 
$\eta$--$\eta^\prime$ system and its $\gamma\gamma$ decays
using a model which is a leading version of the consistently
coupled Schwinger-Dyson (SD) and Bethe-Salpeter (BS) approach.
The electromagnetic interactions are incorporated through
a (generalized) impulse approximation
consistent with this bound--state approach, so
that the Ward-Takahashi identities of QED are preserved
when quarks are dynamically dressed.
To overcome some of the limitations due to the ladder approximation,
we introduce a minimal extension to the bound--state approach employed,
so that the $\mbox{\rm U}_{\rm A}{\rm (1)}$ problem is avoided. 
Pointing out which of our predictions hold in the coupled SD-BS
approach in general, and which are the consequences of the specific,
chosen model, we present the results for the axial-current decay
constants of $\eta_8$, $\eta_0$, and of their physical combinations
$\eta$ and $\eta^\prime$, the results for the $\gamma\gamma$-decay 
constants of $\eta_0$ and $\eta_8$, for the two--photon decay widths 
of $\eta$ and $\eta^\prime$, and for the mixing--independent $R$--ratio
constructed from them. 

%
\end{abstract}

\section{Introduction}
\label{INTRO}

\noindent
A particularly interesting example of the applications of 
Schwinger-Dyson equations to hadronic physics (reviewed in,
{\it e.g.}, \cite{RW,Miransky}), is the approach through 
consistently coupled Schwinger-Dyson (SD) equations for quark
propagators and Bethe-Salpeter (BS) equations for bound states
of quarks. Among various studies of this kind, those of 
Jain and Munczek  \cite{jain91,munczek92,jain93b} are 
judged by many as ``the most extensive and phenomenologically
successful spectroscopic studies in the rainbow--ladder 
approximation" \cite{Burden+Qian+al} and therefore often
chosen \cite{RW,Miransky,Burden+Qian+al,Roberts,Burden+al}
as a representative, paradigmatic example of such studies.
The essence of such a treatment of $q\bar q$ bound
states is the solving of the ladder Schwinger-Dyson (SD) 
equation for the dressed quark propagator $S(q)$, and then
solving in the consistent approximation, with this resulting 
dressed quark propagator and with the same interaction 
kernel, the Bethe-Salpeter (BS) relativistic bound-state
equation for a $q\bar q$ meson. This procedure is crucial
for obtaining the mesons from the light pseudoscalar
octet as Goldstone bosons when the chiral symmetry is 
spontaneously broken. 
Thanks to this,
a coupled SD-BS approach (notably, \cite{jain91,munczek92,jain93b})
can reproduce the correct chiral limit behavior (crucial in
the light sector) simultaneously
with the realistic results for heavy mesons.
In \cite{jain91,munczek92,jain93b}, the interaction kernel
is given by a modeled gluon propagator consisting of: 
{\it a)} the well--known perturbative part, reproducing correctly the 
ultraviolet (UV) asymptotic behavior unambiguously required by 
QCD in its high--energy, perturbative regime, and 
{\it b)} the nonperturbative part, which should describe the 
infrared (IR) behavior. Since the IR behavior of QCD is 
still more or less unknown, this latter, nonperturbative
part of the gluon propagator is modeled. 
In \cite{jain91,munczek92,jain93b}, several forms for this
IR--part have been used and their parameters varied, 
with the outcome that results are
not very sensitive to such variations. Jain and Munczek
\cite{jain91,munczek92,jain93b} have succeeded in reproducing
the leptonic decay constants of pseudoscalar mesons, and,
even more importantly, a very large part of the meson spectrum,
except for such elusive cases as the $\eta$--$\eta^\prime$ 
system. 

Such an up till now successful and reputable 
referent model should be tested further by calculating other
quantities ({\it e.g.}, electromagnetic processes) to see
how well it will do. 
This was our motivation for calculating 
$\pi^0, \eta_c, \eta_b\rightarrow \gamma\gamma$ and 
$\gamma^\star\pi^0 \rightarrow \gamma$ in \cite{KeKl1,Lesna},
and Jain and Munczek's model passed this test very well.
Other applications are also under investigation, and still
many others are possible.
However, for the full assessment
of a model and for getting useful insights in how to improve it,
it is also very interesting to see how it performs at the very 
edges of its applicability. 
Although Jain and Munczek's model is cleverly constructed so 
that it  works well for most pseudoscalar and vector mesons 
below, above, and even on the mass scale of $\eta$ and $\eta^\prime$, 
the limitations of the (``improved" \cite{Miransky} or 
``generalized" \cite{Roberts}) ladder approximation employed 
by the model put the $\eta$--$\eta^\prime$ system on such an 
``applicability edge" of this model -- although not beyond it,
contrary to what a pessimist could have concluded. This will be 
clarified below, where we analyze the $\eta$--$\eta^\prime$ system
and its $\gamma\gamma$ decays in Jain and Munczek's model
\cite{jain91,munczek92,jain93b}, demonstrate the abilities
and limitations of this model, and anticipate in which way it 
can be extended to improve further the description of $\eta$
and $\eta^\prime$.

\section{Solving the consistently coupled SD and BS equations}
\label{QBARQ}

\noindent
Dressed quark propagators $S_f(q)$ for various flavors $f$,
	\begin{equation}
	S_f^{-1}(q)
	=
	A_f(q^2)\slashed{q} - B_f(q^2)~, \qquad (f=u,d,s,...)~,
	\label{quark_propagator}
	\end{equation}
are obtained by solving the SD equation, which in the ladder
approximation ({\it i.e.}, with the true quark-gluon vertex
replaced by the bare one, namely $\gamma^\nu \lambda^j/2$) 
becomes
	\begin{equation}
	S_f^{-1}(p) = \slashp - \widetilde{m}_f - i \,
 g_{\rm st}^{\, 2} \, C_F  \int \frac{d^4k}{(2\pi)^4}
		    \gamma^\mu S_f(k) \gamma^\nu G_{\mu\nu}(p-k)~,
       \label{SD-equation}
	\end{equation}
where $\widetilde{m}_f$ is the bare mass of the quark flavor $f$,
breaking the chiral symmetry explicitly, and $C_F$ is the second Casimir 
invariant of the quark representation, here 4/3 for the case of the 
(halved) Gell-Mann matrices $\lambda^j/2$  ($j=1,...,8$) of SU(3)$_c$.
Neglecting ghosts, the product of the strong coupling constant 
$g_{\rm st}$ and the Landau--gauge gluon propagator can be
approximated by the {\it Ansatz} often described as the 
``Abelian approximation" \cite{MarisRoberts97PRC56}:
	\begin{equation}
 g_{\rm st}^{\, 2} \, C_F  G^{\mu\nu}(k) = G(-k^2)
	\left( g^{\mu\nu} - \frac{k^\mu k^\nu}{k^2} \right)~.
	\label{gluon_propagator}
	\end{equation}

As explained in the Introduction, the function $G$ is given by the sum of 
the known perturbative part $G_{UV}$, and the modeled nonperturbative part
$G_{IR}$: 
	\begin{equation}
	G(Q^2) = G_{UV}(Q^2) + G_{IR}(Q^2)~,\;\;(Q^2=-k^2)~.
\label{factorG}
	\end{equation}
In $G_{UV}$, we employ -- following Ref. \cite{jain93b} -- the 
two--loop asymptotic expression for $\alpha_{\rm st}(Q^2)$:
	\begin{equation}
	G_{UV}(Q^2)
	=
		  4\pi C_F \frac{\alpha_{\rm st}(Q^2)}{Q^2}
	\approx
\frac{{ 4\pi^2 C_F} d}{Q^2 \ln(x_0+\frac{Q^2}{\Lambda_{QCD}^2})}
		{\huge \{}
			1
			+
		 b \, \frac{\ln[\ln(x_0+ \frac{Q^2}{\Lambda_{QCD}^2})]}
			       {\ln(x_0+ \frac{Q^2}{\Lambda_{QCD}^2})}
		{\huge \}}~,
\label{gluon_UV}
	\end{equation}
where $d = 12/(33-2N_f)$, $b = 2\beta_2/\beta_1^2 =
2(19N_f/12 -51/4)/(N_f/3 -11/2)^2$. As in \cite{jain93b}, 
we set the number of flavors $N_f=5$, 
$\Lambda_{QCD}=228\,\mbox{\rm MeV}$, and $x_0=10$.
We adopt the modeled $G_{IR}$, together with its parameters
$a=(0.387\,\GeV)^{-4}$ and $\mu=(0.510\,\GeV)^{-2}$, from Ref. \cite{jain93b}:
	\begin{equation}
	G_{IR}(Q^2)
	=
		   { 4\pi^2 C_F } \,a\,Q^2 e^{-\mu Q^2}.
	\label{gluon_IR}
	\end{equation}

Solving (\ref{SD-equation}) for the propagator functions $A_f(q^2)$ 
and $B_f(q^2)$ also yields the constituent quark masses, defined
(at $q^2 = 0$ for definiteness)
as ${\cal M}_f \equiv B_f(0)/A_f(0)$ for the flavor $f$.

The case $\widetilde{m}_f=0$ corresponds to the chiral limit, where
the current quark mass $m_f=0$, and where the constituent quark mass
stems exclusively from dynamical chiral symmetry breaking (D$\chi$SB) 
\cite{jain91}. For $u$ and $d$-quarks, the chiral limit is a very good 
approximation.  Solving of (\ref{SD-equation}) with 
$\widetilde{m}_u = \widetilde{m}_d = 0$ leads to 
${\cal M}_{ud}=B_u(0)/A_u(0)=B_d(0)/A_d(0)=356$ MeV for 
the gluon propagator (\ref{gluon_propagator})-(\ref{gluon_IR})
with parameters quoted above and used in \cite{jain93b}.

When $\widetilde{m}_f\neq 0$, the SD equation (\ref{SD-equation})
must be regularized by a UV cutoff $\Lambda$ \cite{munczek92,jain93b},
and $\widetilde{m}_f$ is in fact a cutoff--dependent quantity. We 
adopted the parameters of \cite{jain93b}, where (for $\Lambda=134$ GeV) 
the bare mass $\widetilde{m}_f(\Lambda^2)=3.1$ MeV -- chosen to 
ultimately lead to the realistically massive pion  --
yields the light, non-strange isosymmetric constituent quark mass
${\cal M}_{ud}=375$ MeV, just 5\% above its value in the chiral limit.
For $s$--quarks, $\, \widetilde{m}_f(\Lambda^2)$ is $73$ MeV, 
giving us the strange quark constituent mass 
${\cal M}_s\equiv B_s(0)/A_s(0)=610$ MeV \cite{jain93b}.

In the chiral limit, 
solving of (\ref{SD-equation}) with $\widetilde{m}_f=0$ is already 
sufficient to give us the Goldstone pion bound-state vertex{\footnote{In 
(\ref{ChLimSol}), we explicitly included the (elsewhere suppressed) flavor 
factor $\lambda^3/\sqrt{2}$, appropriate for $\pi^0$, to emphasize the 
change of our convention with respect to \cite{KeKl1,Lesna}: we now adopt 
the convention of Jain and Munczek's papers \cite{jain91,munczek92,jain93b} 
for the {\it flavor factors}, but not their conventional color
factor of $1/\sqrt{N_c}$. Hence, we have the additional factor of 
$N_c$ multiplying the integral in Eq. (2.8) of Ref. \cite{munczek92}, 
the formula which otherwise specifies our normalization.}}
$\Gamma_\pi$ that is of zeroth order in the pion momentum $p$,
	\begin{equation}
\Gamma_{\pi^0}(q;p^2\! =\! M_\pi^2\! =\! 0)
=  \frac{\lambda^3}{\sqrt{2}}\, \Gamma_{f\!\bar f}(q;p^2\! =\! 0)_{m_f=0}
 = \frac{\lambda^3}{\sqrt{2}}\, 
     \gamma_5 \, \frac{\sqrt{2}\, B_f(q^2)_{m_f=0}}{f_\pi},    
     \label{ChLimSol}
	\end{equation}
leading \cite{bando94,Roberts} to the famous result (Eqs. (\ref{ChLimAmp}) 
and (\ref{AnomAmpl}) below) for the $\pi^0\to\gamma\gamma$ amplitude
due to the Abelian Adler--Bell--Jackiw (ABJ), or axial, anomaly.  

Of course, for heavier $q\bar q$ composites 
one cannot circumvent solving the BS equation
by invoking the chiral-limit (and the soft-limit, $p^\mu \to 0$)
result (\ref{ChLimSol}).
This is obvious when they contain $c$- or $b$-quarks,
for which the whole concept of the chiral limit is of
course useless even qualitatively. When strange quarks
are present, (\ref{ChLimSol}) can be regarded only as an ``exploratory"
\cite{Burden+al} expression, and it is useful for considering
the chiral limit, since this limit is {\it qualitatively} meaningful for
the $s-$quarks. Nonetheless, we need the {\it quantitative} predictions of 
Jain and Munczek's model for the $s\bar s$ pseudoscalar bound state, which 
is not physical, but enters as the heaviest component in the pseudoscalars 
$\eta$ and $\eta^\prime$, introduced in the next section.

Therefore, we must obtain the bound-state vertex $\Gamma_{s\bar s}$
by explicit solving of
	\begin{equation}
		\Gamma_{s\bar s}(q,p) 
                = i g_{\rm st}^2  C_F
		\int \frac{d^4q^\prime}{(2\pi)^4}
		\gamma^\mu S_s(q^\prime + \frac{p}{2})
      \Gamma_{s\bar s}(q^\prime,p)   S_s(q^\prime - \frac{p}{2})
		\gamma^\nu
		G_{\mu\nu}(q-q^\prime)~,
	\label{BSE}
	\end{equation}
the homogeneous BS equation again in the ladder approximation,
consistently with (\ref{SD-equation}). 
For pseudoscalar ($P$) quarkonia,
the complete decomposition of the BS bound state vertex $\Gamma_P$
in terms of the scalar functions $\Gamma^P_i$ is:
	\begin{equation}
\Gamma_P(q,p)= 
	\gamma_5 \left\{\, \Gamma^P_0(q,p) + \slashp \, \Gamma^P_1(q,p)
      + \slashed{q} \, \Gamma^P_2(q,p) +
	[\slashp,\slashed{q}]\, \Gamma^P_3(q,p)\, \right\}.
\label{Decomposition}
	\end{equation}  
(The flavor structure is suppressed again. For neutral pseudoscalars, 
$\Gamma_P$ is decomposed into $f\bar f$-components $\Gamma_{f\!\bar f}$ 
according to Eq. (\ref{neutralP}) below.) The BS equation (\ref{BSE}) 
leads to a coupled set of integral equations for the functions 
$\Gamma^P_i$ $(i=0,...,3)$, which we find to be most easily solved 
numerically in the Euclidean space by following the procedure of 
Jain and Munczek \cite{jain91,munczek92,jain93b}, who formulate 
the problem in terms of the BS amplitudes $\chi_{f\!\bar f}(q,p) 
\equiv S_f(q + {p}/{2}) \Gamma_{f\! \bar f}(q,p) S_f(q - {p}/{2})$.

In order to avoid the angular integration,
we also adopt the momentum expansion (in the Chebyshev polynomials)  
\cite{jain91,munczek92,jain93b} of the four scalar functions 
appearing in the decomposition of the BS amplitudes.   
Ref. \cite{jain93b} often kept 
only the lowest order moment in the 
Chebyshev expansion,
because they found it adequate for most of the meson spectrum. 
In contrast, 
while using the kernel and parameters of Ref. \cite{jain93b},
we always retain all four functions in solving of the BS
equation (\ref{BSE}), and the first two moments
in the Chebyshev expansion. The accuracy of this procedure has 
recently received an independent confirmation  
-- especially for presently interesting charge conjugation eigenstates --
from Maris and Roberts \cite{MarisRoberts97PRC56}. 
In their study of $\pi$- and $K$-meson BS amplitudes, 
they employed {\it both} the Chebyshev decomposition, 
and straightforward multidimensional integration. Their comparison 
of these two techniques showed the very quick convergence of the 
Chebyshev expansion: in the case of equal quark and antiquark masses, 
such as in the pion, the zeroth and the first Chebyshev moment are enough 
for an accurate representation of the solution. Even for the kaon, still 
just one more is needed \cite{MarisRoberts97PRC56}, in spite of the 
difference in the masses of its constituents. 
(Of course, the limitations of the ladder approximation 
would in the end lead to increasing difficulties 
if one of the fermion messes became much larger still, 
as recognized also by \cite{munczek92}. However, 
if the mass ratio of the constituents is not too large, various 
contributions beyond ladder approximation largely cancel out in 
the flavor-nonsinglet pseudoscalar, vector and axial channels 
\cite{BenderRobertsSmekal96,RobertsNT9609039,MarisRoberts97PRC56},
explaining the success of the ladder approximation in these channels.)

Our procedure, already successfully used in \cite{KeKl1,Lesna} for
$M_{\eta_c}$ and $M_{\eta_b}$, gives us $M_{s\bar s} = 721$ MeV
for the unphysical pseudoscalar $s\bar s$ bound state entering 
in the $\eta$--$\eta^\prime$ system in the fashion discussed 
in the next section. Naturally, when we abandon the chiral
limit approximation in (\ref{SD-equation}), we can also obtain 
the (isosymmetric) pion bound-state vertex 
$\Gamma_{\pi^0}=\Gamma_{u\bar u}=\Gamma_{d\bar d}$,
replacing $s \to u$ in (\ref{BSE}). Although we stress that
the chiral limit is an excellent approximation for many purposes
in the case of pions, including the computation of $\pi^0 \to
\gamma\gamma$, it is also very important that the experimental
$\pi^0$-mass $M_{\pi^0}=135$ MeV is reproduced \cite{jain93b} 
through (\ref{BSE}) as $M_{u\bar u}$ (= $M_{d\bar d}$) with
the small explicit chiral symmetry breaking, 
${\widetilde m}_{ud}(\Lambda^2)= 3.1$ MeV, 
corresponding to (isosymmetric) current $u$- and $d$-quark masses 
$m=8.73$ MeV, close to the empirical values extracted by current algebra.
Such a small $m$ cannot jeopardize the relevance of
(\ref{ChLimSol}) for the computation of $\pi^0\to\gamma\gamma$,
as shown also by Ref. \cite{Frank+al}, which found
(in an approach closely related to ours) that the
amplitude decreased with respect to the analytic, chiral-limit
axial anomaly result only by less than 1\% when they
introduced the non-vanishing but small $u,d$-quark mass
$m=6.7$ MeV.

\section{ $\eta$--$\eta^\prime$ complex and its
axial-current decay constants}
\label{AXIAL}

\noindent
The SU(3)$_f$ octet and singlet isospin zero states,
$\eta_8$ and $\eta_0$, are in the $q\bar q$-basis given by
	\begin{eqnarray}
	|\eta_8\rangle
	&=&
	\frac{1}{\sqrt{6}} (|u\bar{u}\rangle + |d\bar{d}\rangle
					    -2 |s\bar{s}\rangle)~,
\label{eta8def}
	\\
	|\eta_0\rangle
	&=&
	\frac{1}{\sqrt{3}} (|u\bar{u}\rangle + |d\bar{d}\rangle
					     + |s\bar{s}\rangle)~.
\label{eta0def}
	\end{eqnarray}
In our phenomenologically successful model choice \cite{jain93b},
the flavor SU(3)$_f$-symmetry is broken by the $s$-quark
mass being realistically larger than the $u,d$-masses.
Nevertheless, the isospin symmetry for $u$ and $d$-quarks is assumed
exact throughout this paper. As is most commonly done, 
Eqs. (\ref{eta8def}) and (\ref{eta0def}) both
employ the same quark basis states $|f\bar{f}\rangle$ ($f=u,d,s$)
to define $\eta_8$ and $\eta_0$. As pointed out by Gilman and
Kauffman \cite{GilKauf} (following Chanowitz, their Ref. [8]),
this usual procedure implicitly assumes nonet symmetry. However, 
it is ultimately broken by nonabelian (``gluon") axial anomaly,
which will be discussed in Sec.~\ref{ETA-ETAPRIME}.

$\eta_8$ and $\eta_0$ cannot be physical as they are not the mass 
eigenstates; that are their mixtures $\eta$ and $\eta^\prime$:
	\begin{eqnarray}
	|\eta\rangle &=& \cos\theta\, |\eta_8\rangle
		       - \sin\theta\, |\eta_0\rangle~,
\label{etadef}
	\\
	|\eta^\prime\rangle &=& \sin\theta\, |\eta_8\rangle
			      + \cos\theta\, |\eta_0\rangle~.
\label{etaPrimedef}
	\end{eqnarray}
The determination of the specific value that the mixing angle $\theta$ 
should take, is a difficult issue which will be handled separately
in Sec.~\ref{ETA-ETAPRIME}. 
We will keep our discussion general till we
evaluate those of our results which are independent of
the mixing and $\theta$ -- such as the decay constants of
of the unmixed states $\eta_8$ and $\eta_0$ -- and point out 
those quantities for evaluation of which we need a concrete 
value of $\theta$. 

For the light neutral pseudoscalar mesons $P=\pi^0,\eta_8,\eta_0$, 
their axial-current decay constants 
$f_P = f_{\pi^0}$, $f_{\eta_8}$ and $f_{\eta_0}$, 
are defined by the matrix elements
\begin{equation}
\langle 0|\bar{\psi}(0) \gamma^\mu \gamma_5 \frac{\lambda^j}{2} \psi(0) 
|P(p) \rangle = i \, \delta^{jP}  f_{P} \,  p^\mu  \, ,
\label{deffax}
\end{equation}
where $\psi=(u,d,s)$ is the fundamental representation of SU(3)$_f$, 
while $P=\pi^0,\eta_8,\eta_0$ simultaneously has the meaning of the
respective SU(3)$_f$ indices 3,8,0. This picks out the diagonal 
($j=3,8$) SU(3)$_f$ Gell-Mann matrices $\lambda^j$, and 
$\lambda^0 \equiv (\sqrt{2/3}){\bf 1}_3$, in Eq. (\ref{deffax}).

The neutral pseudoscalars $P$ are expressed through the
quark basis states $|f\bar{f}\rangle$ by
\begin{equation}
| P \rangle = \sum_f \, \left(\frac{\lambda^P}{\sqrt{2}}\right)_{f\! f} 
	      \, |f\bar f \rangle \, 
\equiv \sum_f \,  a_f^P \, |f\bar f \rangle \, , \qquad (f = u, d, s) \, ,  
\label{neutralP}
\end{equation}
where the nonvanishing coefficients 
$a^P_f \equiv (\lambda^P/\sqrt{2})_{f\! f}$ for $P=\pi^0$ are 
$a_u^{\pi^0}=-a_d^{\pi^0}=1/\sqrt{2}=(\lambda^3/\sqrt{2})_{11}$,
whereas for $\eta_8$ they are  $a_u^{\eta_8} = a_d^{\eta_8} =
1/\sqrt{6} = (\lambda^8/\sqrt{2})_{11}= (\lambda^8/\sqrt{2})_{22}$,
$a_s^{\eta_8} = - 2/\sqrt{6}=(\lambda^8/\sqrt{2})_{33}$, and for 
$\eta^0$, $a_u^{\eta_0} = a_d^{\eta_0} = a_s^{\eta_0} = 
(\lambda^0/\sqrt{2})_{ff}=1/\sqrt{3}$.

The axial-current decay constants defined in (\ref{deffax}) can 
be expressed as 
\begin{equation}
f_{P} = \sum_{f=u,d,s} \, 
\frac{({\lambda^P}_{\!f\!f})^2}{2}\, f_{f\!\bar f} \, ,\qquad 
(\, P=\pi^0\leftrightarrow 3, \eta_8\leftrightarrow 8,
\mbox{\rm and}\,\,\, \eta_0\leftrightarrow 0 \,)\, ,
\label{axfP}
\end{equation}
where we have for convenience introduced the auxiliary decay constant
$f_{f\!\bar f}$, defined as the decay constant of the $f\bar f$-pseudoscalar 
bound state which has the mass ${M_{f\!\bar f}}$ and is described by the
BS vertex $\Gamma_{f\!\bar f}(q,p)$, so that using the definitions of
Bethe-Salpeter bound-state amplitudes or vertices in the matrix elements
(\ref{deffax}) as in, {\it e.g.},
\cite{jain91,munczek92,jain93b,Burden+Qian+al}, leads to
	\begin{equation}
	f_{f\!\bar f} = i \frac{N_c}{\sqrt{2}}
		\frac{1}{{M_{f\!\bar f}}^2}
		\int \frac{d^4q}{(2\pi)^4}
		\mbox{\rm tr} \left[ \, \slashp \, \gamma_5 
S_f(q+\frac{p}{2}) \Gamma_{f\!\bar f}(q,p) S_f(q-\frac{p}{2}) \right] \, .
\label{f_ff}
	\end{equation}
It turns out that this equation can also be applied for $M_{f\!\bar f} = 0$,
as the limit exists. 

In the isospin limit, we get
$f_{\pi^0}\equiv f_{u\bar u}=f_{d\bar d}= f_{u\bar d}\equiv
f_{\pi}=93.2$ MeV in our chosen model \cite{jain93b}.
For the axial decay constant of the $s\bar s$ pseudoscalar bound state,
we obtain  $f_{s\bar s}=136.5 \MeV = 1.47 f_\pi$. (This factor with respect
to $f_\pi$ is very reasonable and even expected, since the model 
\cite{jain93b} also predicts the decay constant of the charged kaon 
$f_{K^+}=f_{u\bar d}=114 \, \MeV = 1.23 f_\pi$.) Eq. (\ref{axfP}) then 
yields $f_{\eta_8}=122.1$ MeV and $f_{\eta_0}=107.6$ MeV. Note that 
$f_{\eta_8}=1.31 \, f_{\pi}$, which is rather close to the result 
$f_{\eta_8}=1.25 \, f_{\pi}$ \cite{DonoghueHolsteinLin}, obtained 
in the chiral perturbation theory ($\chi$PT).

Evaluating the matrix elements of the pertinent 
mixtures yields the $\eta$ and $\eta^\prime$ decay constants 
\begin{equation}
f_{\eta} =  \left( \frac{1}{\sqrt{3}}\cos\theta - 
\frac{\sqrt{2}}{\sqrt{3}} \sin\theta  \right)^2 f_\pi 
+  \left( - \frac{\sqrt{2}}{\sqrt{3}} \cos\theta - 
\frac{1}{\sqrt{3}} \sin\theta \right)^2 f_{s\bar s} \, ,
\label{feta}
\end{equation}
\begin{equation}
f_{\eta^\prime} =  \left( \frac{\sqrt{2}}{\sqrt{3}} \cos\theta + 
\frac{1}{\sqrt{3}} \sin\theta \right)^2 f_\pi
+ \left( \frac{1}{\sqrt{3}}\cos\theta - 
\frac{\sqrt{2}}{\sqrt{3}} \sin\theta  \right)^2 f_{s\bar s} \, .
\label{fetaprime}          
\end{equation}

\section{
$\pi^0, \eta_8, \eta_0 \to \gamma\gamma$ and 
$\eta, \eta^\prime \rightarrow \gamma\gamma$ processes}
\label{2GAMMA}

\noindent
The transition amplitudes for $\eta, \eta^\prime \rightarrow \gamma\gamma$ 
can be obtained from the $\gamma\gamma$-transition amplitudes for
$\eta_8$ and $\eta_0$ by forming the  appropriate mixtures, in line with
(\ref{eta8def})-(\ref{etaPrimedef}). 
The $\eta_8,\eta_0\to\gamma\gamma$ amplitudes are in turn calculated in
the same way as $\pi^0, \eta_c, \eta_b \rightarrow \gamma\gamma$ 
in \cite{KeKl1,Lesna}.

This means that we assume that these decays proceed through the
triangle graph (depicted in Fig. 1), and that we calculate 
the pertinent amplitudes \cite{itzykson80}
\begin{equation}
T_P^{\mu\nu}(k,k^\prime)
      = 
	\varepsilon^{\alpha\beta\mu\nu} k_\alpha k^\prime_\beta
	T_P(k^2,k^{\prime 2})~,
\end{equation}
\noindent
and the corresponding on-shell ($k^2=0$ and $k^{\prime 2}=0$) 
decay widths 
\begin{equation}
	       W(P\to\gamma\gamma)
	= 
	{\frac{\pi\alpha_{\rm em}^2}{4}} M_P^3
     \, |T_P(0,0)|^2~ \, ,  \qquad
(P = \pi^0, \eta, \eta^\prime, ... ) \, ,
\label{PdcyWdth}
\end{equation}
using the framework advocated by (for example) 
\cite{bando94,Roberts,Frank+al,Burden+al,AlkRob96}
in the context of electromagnetic interactions of BS bound states,
and often called the generalized impulse approximation (GIA) -
{\it e.g.}, by \cite{Frank+al,Burden+al}.
To evaluate the triangle graph, we therefore use the
{\it dressed} quark propagator $S_f(q)$, Eq.~(\ref{quark_propagator}),
and the pseudoscalar BS bound--state vertex $\Gamma_P(q,p)$
instead of the bare $\gamma_5$ vertex.
Another ingredient, crucial for GIA's ability to
reproduce the correct Abelian anomaly result, is
employing an appropriately dressed
{\it electromagnetic} vertex $\Gamma^\mu_f(q^\prime,q)$,
which satisfies the vector Ward--Takahashi identity (WTI),
	\begin{equation}
	(q^\prime-q)_\mu \Gamma^\mu_f(q^\prime,q) =
		S^{-1}_f(q^\prime) - S^{-1}_f(q)~,
	\qquad (f = u, d, s, ... ).
	\label{WTI-v}
	\end{equation}
\noindent Namely, assuming that photons
couple to quarks through the bare vertex $\gamma^\mu$
would be inconsistent with
our quark propagator, which, dynamically dressed through
Eq.~(\ref{SD-equation}), contains the momentum-dependent
functions $A_f(q^2)$ and $B_f(q^2)$.
The bare vertex $\gamma^\mu$ obviously violates (\ref{WTI-v}),
implying the nonconservation of the electromagnetic current
and of the electric charge. 
Since solving the pertinent SD equation for the  
dressed quark-photon vertex $\Gamma^\mu_f$ is a difficult 
problem that has only recently begun to be addressed \cite{F}, 
it is customary to use realistic {\it Ans\"{a}tze}. 
Following, {\it e.g.}, \cite{Frank+al,Burden+al,Roberts,AlkRob96}, we 
choose the Ball--Chiu ~\cite{BC} vertex:
	\begin{eqnarray}
	\Gamma^\mu_f(q^\prime,q) =
	A_{\bf +}^f(q^{\prime 2},q^2)
       \frac{\gamma^\mu}{\textstyle 2}
	+ \frac{\textstyle (q^\prime+q)^\mu }
	       {\textstyle (q^{\prime 2} - q^2) }
	\{A_{\bf -}^f(q^{\prime 2},q^2)
	\frac{\textstyle (\slashed{q}^\prime + \slashed{q}) }{\textstyle 2}
	 - B_{\bf -}^f(q^{\prime 2},q^2) \}~,
	\label{BC-vertex}
	\end{eqnarray}
where
$H_{\bf \pm}^f(q^{\prime 2},q^2)\equiv [H_f(q^{\prime 2})\pm H_f(q^2)]$,
for $H = A$ or $B$. 
This {\it Ansatz}:
{\it - i)} satisfies the WTI (\ref{WTI-v}),
{\it - ii)} reduces to the bare vertex in the free-field limit
as must be in perturbation theory,
{\it - iii)} has the same transformation properties under
Lorentz transformations and charge conjugation as the
bare vertex,
{\it - iv)} has no kinematic singularities, and 
{\it - v)} does not introduce any new parameters as it is completely
determined by the quark propagator (\ref{quark_propagator}).

For the meson $P$ whose flavor content is given by Eq. (\ref{neutralP}), 
GIA yields the amplitude
\begin{displaymath}
	T_{P}^{\mu\nu}(k,k^\prime)
	=
	\sum_{f=u,d,s} \, a_f^P \, Q_f^2 \, 
	N_c \, (-)
	\int\frac{d^4q}{(2\pi)^4} \mbox{\rm tr} \{
	\Gamma^\mu_f(q-\frac{p}{2},k+q-\frac{p}{2})
	S_f(k+q-\frac{p}{2})
\end{displaymath}
\begin{equation}
	  \qquad
	\times
	\Gamma_f^\nu(k+q-\frac{p}{2},q+\frac{p}{2})
	S_f(q+\frac{p}{2})
	\Gamma_{f\bar f}(q,p)
	S_f(q-\frac{p}{2}) \}
	+
	(k\leftrightarrow k^\prime,\mu\leftrightarrow\nu).
\label{Tmunu(2)}
\end{equation}
The coefficients $a^P_f$ of various flavor components 
$|f\bar f \rangle$ in $P=\pi^0, \eta_8, \eta_0$, are given below 
Eq. ~(\ref{neutralP}).
$Q_f$ denotes the charge of the quark flavor $f$.
The dependence on the flavor $f$ has been indicated
on the BS vertices, dressed propagators and
electromagnetic vertices in the loop integral for each
quark flavor. It is convenient to 
separate out the $a^P_f$ and $Q_f^2$ dependence by
denoting each integral (times $( - N_c)$) in (\ref{Tmunu(2)}) 
the ``reduced $\gamma\gamma$-amplitude" ${\widetilde T}^{\mu\nu}_{f\bar f}$. 
The ``reduced scalar amplitude" ${\widetilde T}_{f\bar f}$ 
for the flavor $f$ is then
\begin{equation}
{\widetilde T}_{f\bar f}^{\mu\nu}(k,k^\prime)
	=
	\varepsilon^{\alpha\beta\mu\nu} k_\alpha k^\prime_\beta
       {\widetilde T}_{f\bar f}(k^2,k^{\prime 2})~.
\label{ReducScal}
\end{equation}

{\subsection{$\gamma\gamma$-amplitudes and 
                                  $\gamma\gamma$-decay constants}}

\noindent Regardless of what the chiral-limit
solutions for the propagator (\ref{quark_propagator}) and
the bound-state vertex (\ref{ChLimSol}) are in detail,
${\widetilde T}_{f\bar f}^{\mu\nu}(0,0)$ can be evaluated 
analytically in the chiral (and soft) limit \cite{Roberts,bando94}, 
which is perfectly adequate for $f=u,d$, {\it i.e.}, for a Goldstone 
$P=\pi^0$. There,
\begin{equation}
{\widetilde T}_{\pi^0}(0,0) \equiv
{\widetilde T}_{u\bar u}(0,0)={\widetilde T}_{d\bar d}(0,0)
= \frac{N_c}{2\sqrt{2}\pi^2 f_\pi}  \, ,
\label{ChLimAmp}
\end{equation}
to which we stick throughout. In terms of  
\begin{equation}
T_{P}(k^2,k^{\prime 2}) \equiv
 \sum_f \, a_f^{P} \, Q_f^2 \, 
{\widetilde T}_{f\bar f}(k^2,k^{\prime 2}) \, ,
\label{DefReducScal}
\end{equation}
this leads to the standard form of the successful
axial-anomaly-result{\footnote{We can also get 
-- in the fashion of Ref. \cite{AlkRob96} -- 
the anomalous ``box'' amplitude for $\gamma^\star\to\pi\pi\pi$.}
for $\pi^0\to\gamma\gamma$:
\begin{equation}
 T_{\pi^0}(0,0) = 
\frac{N_c}{2\sqrt{2}\pi^2 f_\pi}\, \sum_f \, a_f^{\pi^0} \, Q_f^2 \,
= \frac{1}{4\pi^2 f_\pi} \, .
\label{AnomAmpl}
\end{equation}
Note that this reproduction of the chiral limit relation between the 
$\pi^0\to\gamma\gamma$ decay amplitude and the pion axial-current
decay constant $f_\pi$, is not dependent on the pion's internal 
structure (or the interaction kernel that produces it) in any way
\cite{Roberts,bando94} -- and this is an important advantage of the 
coupled SD-BS approach over most other bound-state approaches, since
the axial anomaly is on fundamental grounds known to be independent 
of the structure. 
(Those calculations of $\pi^0\to\gamma\gamma$ which rely on the
details of the hadronic structure -- be it in the context of the
BS equation without D$\chi$SB, nonrelativistic quarks, or otherwise --
have problems to describe this decay accurately even when the model
parameters are fine-tuned for that purpose; {\it e.g.}, see 
\cite{HorbKoniuk93,GuiasuKoniuk93,Munz+al94,Munz96} and references 
therein. The most successful of these model fits, Ref.\cite{Munz+al94}, 
numerically obtains the width of 7.6 eV at the expense of
fine-tuning constituent quark masses to unusually small values.)
Of course, $f_\pi$ itself {\it is} structure dependent. It is a 
{\it calculated quantity} in the SD-BS approach. Our model choice 
\cite{jain93b} successfully reproduces the experimental value of 
$f_\pi$, and this is obviously of utmost importance for the 
theoretical description of anomalous processes.

The implications thereof for the $\eta_8$, $\eta_0$ and their 
mixtures $\eta$ and $\eta^\prime$ are now clear, because
those parts of their $\gamma\gamma$-decay amplitudes which
stem from their $u\bar u$ and $d\bar d$ components are
(just as in $\pi^0$) accurately given by the Abelian anomaly
({\it i.e.}, Eq. (\ref{ChLimAmp})) for {\it any} interaction kernel 
which leads to the correct $f_\pi$ 
-- be it the present one, or some improved one.  

In other words, Eq. (\ref{ChLimAmp}) implies that the uncertainty 
(in the $\gamma\gamma$-amplitudes) due to modeling of the interaction 
kernel and the resulting bound state, is to a large extent cornered 
into the $s\bar s$ sector, since only ${\widetilde T}_{s\bar s}(0,0)$ 
-- the $\gamma\gamma$-decay amplitude of the relatively heavy 
$s\bar s$-pseudoscalar -- has to be evaluated numerically. 
{}From (\ref{Tmunu(2)})--(\ref{ReducScal}), we find numerically 
that in the model of \cite{jain93b}, 
${\widetilde T}_{s\bar s}(0,0)=0.62\, {\widetilde T}_{u\bar u}(0,0)$. 

The $\pi^0\to\gamma\gamma$ decay amplitude $T_{\pi^0}(0,0)$ 
at any pion mass can be used as a definition of pionic 
$\gamma\gamma$-decay constant ${\bar f}_\pi$ through
$T_{\pi^0}(0,0)\equiv 1/4\pi^2 {\bar f}_\pi =
({N_c}/{2\sqrt{2}\pi^2 {\bar f}_\pi})\, \sum_f \, a_f^{\pi^0} \, Q_f^2 \,$. 
Eq. (\ref{AnomAmpl})
then reveals that ${\bar f}_\pi = f_\pi$ in the chiral limit,
which result is well-known from the axial anomaly analysis.
Although the chiral limit formula (\ref{AnomAmpl}) can be applied 
without reservations only to pions, it is for historical reasons 
customary to write the amplitudes for $\eta_8, \eta_0 \to \gamma\gamma$ 
in the same form as (\ref{AnomAmpl}), defining thereby the 
$\gamma\gamma$-decay constants ${\bar f}_{\eta_8}$ and ${\bar f}_{\eta_0}$: 
\begin{equation}
 T_{\eta_8}(0,0) \equiv
\frac{N_c}{2\sqrt{2}\pi^2 {\bar f}_{{\eta_8}}}\, 
\sum_f \, a_f^{\eta_8} \, Q_f^2 \, =
\frac{f_\pi}{{\bar f}_{{\eta_8}}}\, \frac{T_{\pi^0}(0,0)}{\sqrt{3}} ,
\label{eta8Ampl}
\end{equation}
\begin{equation}
 T_{\eta_0}(0,0) \equiv 
\frac{N_c}{2\sqrt{2}\pi^2 {\bar f}_{\eta_0}}\, 
\sum_f \, a_f^{\eta_0} \, Q_f^2 \, =
\frac{f_\pi}{{\bar f}_{{\eta_0}}}\, 
\frac{\sqrt{8}\, T_{\pi^0}(0,0)}{\sqrt{3}} .
\label{eta0Ampl}
\end{equation}
As pointed out by \cite{Donoghue+alKnjiga}, 
${\bar f}_{\eta_8}$ and ${\bar f}_{\eta_0}$
are {\bf not} {\it a priori} simply
connected with the usual axial-current decay constants $f_{\eta_8}$
and $f_{\eta_0}$, in contradistinction to the pion case, where
$f_\pi = {\bar f}_\pi$ because the chiral limit is such a good 
approximation for pions.

Eqs. (\ref{DefReducScal})-(\ref{eta0Ampl}) reveal that
in the present approach ${\bar f}_{\eta_8}$ and ${\bar f}_{\eta_0}$
are naturally expressed through $f_\pi$ ({\it i.e.}, through
${\widetilde T}_{u\bar u}(0,0) =  {\widetilde T}_{d\bar d}(0,0)$
evaluated in the chiral limit), and ${\widetilde T}_{s\bar s}(0,0)$,
the $\gamma\gamma$--decay amplitude of the unphysical pseudoscalar
$s\bar{s}$ bound state, calculated for nonvanishing $m_s$.
Our predictions for ${\bar f}_{\eta_8}$ and ${\bar f}_{\eta_0}$
are thus:
\begin{equation}
{\bar f}_{\eta_8} = \frac{3 \, f_\pi}{5 - 4\pi^2\sqrt{2}f_\pi  
				{\widetilde T}_{s\bar s}(0,0)/N_c} \, ,
\label{f8}
\end{equation}
\begin{equation}
{\bar f}_{\eta_0} = \frac{6 \, f_\pi}{5 + 2\pi^2 \sqrt{2} f_\pi
				{\widetilde T}_{s\bar s}(0,0)/N_c} \, .
\label{f0}
\end{equation}
Derivation of Eqs. (\ref{f8}) and (\ref{f0}) shows that irrespective 
of any specific model choice, any $q\bar q$ bound-state approach 
(such as our coupled SD-BS approach in conjunction with GIA) 
{\it which has the merit} of reproducing the anomalous
$\pi^0\to\gamma\gamma$ amplitude in the chiral limit, Eq. (\ref{ChLimAmp}) 
or (\ref{AnomAmpl}), should give the relations (\ref{f8}) and (\ref{f0})
for ${\bar f}_{\eta_8}$ and ${\bar f}_{\eta_0}$, when pions are 
approximated by the chiral limit. The concrete numerical values of
${\bar f}_{\eta_8}$ and ${\bar f}_{\eta_0}$ depend on what are the 
model predictions for $f_\pi$ and ${\widetilde T}_{s\bar s}$.

Since in the coupled SD-BS approach we can numerically evaluate 
${\widetilde T}_{s\bar s}(0,0)$ for arbitrary values of the $s$-quark 
mass, Eqs. (\ref{f8}) and (\ref{f0}) give our predictions for the 
effects of the $SU(3)_f$ breaking on $\gamma\gamma$-decays in the 
$\eta$--$\eta^\prime$ system. In the $SU(3)_f$ limit 
(where ${\widetilde T}_{s\bar s} = {\widetilde T}_{u\bar u}$) 
and the chiral limit applied also to $s$-quarks,
we obviously recover ${\bar f}_{\eta_8} = f_\pi$, but also 
${\bar f}_{\eta_0} = f_\pi$, since nonet symmetry in the sense 
of \cite{GilKauf} is a starting assumption of ours.

For our present model choice \cite{jain93b}, where
${\widetilde T}_{s\bar s}(0,0)=0.62\, {\widetilde T}_{u\bar u}(0,0)$,
Eqs. (\ref{f8}) and (\ref{f0}) give
        \begin{eqnarray}
        \bar{f}_{\eta_8} &=& 73.64~\MeV = 0.797 f_\pi~,
\label{f8value}
        \\
        \bar{f}_{\eta_0} &=& 98.58~\MeV = 1.067 f_\pi~.
\label{f0value}
        \end{eqnarray}
While this $\bar{f}_{\eta_0}$ agrees with the results of $\chi$PT
\cite{DonoghueHolsteinLin,BBC90}, there is a difference concerning
$\bar{f}_{\eta_8}$, since ${\bar f}_{\eta_8} > f_\pi$ in $\chi$PT.
This is important because the value of $f_\pi/{\bar f}_{\eta_8}$ has 
impact on the possible values of the $\eta$--$\eta^\prime$ mixing 
angle. We therefore devote the following subsection to the discussion
of this result and the meaning of this difference.

{\subsection{In SD-BS approach, ${\bar f}_{\eta_8} < f_\pi$ generally}}
\label{discussf8}

\noindent The ${\bar f}_{\eta_8}$-value (\ref{f8value}) 
is a result of a specific model. 
However, for the $s$-quark mass realistically heavier
than the $u,d$-quark masses, ${\bar f}_{\eta_8} < f_\pi$ 
holds in the coupled SD-BS approach generally, {\it i.e.},
independently of chosen model details.
To see this, let us start by noting that 
${\bar f}_{\eta_8} < f_\pi$ is  equivalent to 
$T_{\eta_8}(0,0) > T_{\pi^0}(0,0)/\sqrt{3}$,
and since we can re-write Eq. (\ref{DefReducScal}) for $\eta_8$ as
\begin{equation}
T_{\eta_8}(0,0)=\frac{T_{\pi^0}(0,0)}{\sqrt{3}}+\frac{1}{9}
\frac{2}{\sqrt{6}}
\left[{\widetilde T}_{d\bar d}(0,0)-{\widetilde T}_{s\bar s}(0,0)\right]\, ,
\label{rewrite}
\end{equation}
the inequality ${\bar f}_{\eta_8} < f_\pi$ is in our approach 
simply the consequence of the fact that 
the (``reduced") $\gamma\gamma$-amplitude of the $s\bar s$-pseudoscalar 
bound state, ${\widetilde T}_{s\bar s}$,
is smaller than the corresponding 
non-strange $\gamma\gamma$-amplitude ${\widetilde T}_{d\bar d}$ 
($={\widetilde T}_{u\bar u}={\widetilde T}_{\pi^0}$ in the 
isosymmetric limit), for any realistic relationship between 
the non-strange and much larger strange quark masses.

Only in the chiral limit (and close to it), subtle  cancellations
between the bound-state vertices, WTI-preserving $qq\gamma$ vertices
and dynamically dressed propagators lead to the large anomalous
amplitude (\ref{ChLimAmp}), or its slight modification (the size of
which is controlled by Veltman-Sutherland theorem) for small $u$ and $d$ 
masses. Significantly away from the chiral limit, 
what happens is basically simple suppression of
${\widetilde T}_{f\bar f}(0,0)$ by the large quark mass in the 
propagators in the triangle loop of Fig. 1. 
Essentials and generality of the suppression mechanism can be understood 
in basic terms in two (related) ways, through the 
{\it simple free quark loop} (QL) model and 
the Goldberger-Treiman (GT) relation. 

{\it i)} In a QL model ({\it e.g.}, see \cite{Ametller+al92} and 
references therein),  the strength of the Yukawa point couplings 
of the free quarks of the flavor $f$ to the pseudoscalar $P$ is 
given by the constant $g_f$, and quarks have constant constituent 
masses ${\cal M}_f$ (in contradistinction to the momentum-dependent 
mass functions ${\cal M}_f(q^2) = B_f(q^2)/A_f(q^2)$ in our framework). 
Up to some arcsine-type dependence unessential here, each flavor $f$ 
then contributes simply $(g_{P\!f\!f}/{\cal M}_f)\, Q_f^2 \equiv 
(g_f/{\cal M}_f) \, a_f^P \, Q_f^2$ to the triangle-loop 
$\gamma\gamma$-amplitude \cite{Ametller+al92}. 
In the case of the strictly SU(3)$_f$-symmetric coupling, the
Yukawa couplings would be the same for all flavors, $g_f = g$.
The broken SU(3)$_f$-symmetry implies that $g_f$ {\it can}  
differ for various flavors $f$ -- but not by much, so that relative
strengths of the factors $g_f/{\cal M}_f$ for various flavors is 
essentially determined by $1/{\cal M}_f$.
Actually, this is what we find in our SD-BS framework, where
the pseudoscalar bound-state vertices
$\Gamma_{f\bar f}$ are analogous to the coupling 
$g_f$ in the QL model, and $g_f/{\cal M}_f$ is analogous to 
our ``reduced" amplitude ${\widetilde T}_{f\bar f}(0,0)$.
Obviously, our approach allows for the flavor dependence of our BS 
$Pq\bar q$ vertices $\Gamma_{f\!\bar f}$, but due to the fact that the
broken SU(3)$_f$ is still an approximate symmetry, their variation 
with the breaking, given in terms of strange-to-nonstrange 
constituent mass ratio, is rather weak and cannot influence much
the suppression occurring as the constituent mass in the denominator 
grows significantly. Hence, essentially the same mechanism is at work as 
in the QL model. That this parallel works very well, can be seen 
from the fact that the inverse of the strange-to-nonstrange
constituent mass ratio in our SD-BS model, namely 1.63, quite
accurately reproduces the suppression of the $s\bar s$ decay 
amplitude  
${\widetilde T}_{s\bar s}(0,0) = 0.62\, {\widetilde T}_{u\bar u}(0,0)$,
found numerically from (\ref{Tmunu(2)})--(\ref{ReducScal}).

{\it ii)} A related way to see the same effect is to apply the 
quark--level GT relation, $g_f/{\cal M}_f = 1/f_{f\!\bar f}$, 
to the pseudoscalars with the $f\bar f$ quark content in the 
QL model. Then, roughly the same suppression factor occurs again, 
due to $f_{s\bar s} = 1.47\, f_\pi$. 
This is only roughly, since $s\bar s$ is further away from the 
chiral limit than $u\bar u$ and $d\bar d$ constituting the pion.
Nevertheless, invoking the GT relation is in fact a very robust 
way to show that  
${\widetilde T}_{s\bar s}(0,0) < {\widetilde T}_{u\bar u}(0,0)$
must surely hold, even though $s$-quarks are much lighter than
$c$- or $b$-quarks (where the suppression is by orders of magnitude
\cite{KeKl1}), and D$\chi$SB is for $s$-quarks of importance  
similar to that for $u,d$-quarks. Precisely because the chiral limit
makes sense for $s$-quarks {\it qualitatively} (as the pseudoscalars
containing $s$-quarks, can still be considered pseudo-Goldstone
bosons), the GT relation must continue to hold {\it approximately}
in the $s\bar s$-sector, regardless of any specific interaction kernel 
and of the resulting hadronic structure. 
${\widetilde T}_{s\bar s}(0,0) < {\widetilde T}_{u\bar u}(0,0)$ is
therefore obligatory simply due to $f_{s\bar s} > f_\pi$. 

The GT relation is useful also for demonstrating the robustness even 
of our model-dependent result on ${\widetilde T}_{s\bar s}(0,0)$ -- 
namely, that in spite of the model-dependence in the $s$-quark sector, 
the model kernel of our choice \cite{jain93b} should not lead to 
$\gamma\gamma$-amplitude ${\widetilde T}_{s\bar s}(0,0)$ 
excessively different from the ones which would result from
an improved kernel. 
The usage of the GT relation at the quark level is especially transparent
in the context of the simple free quark loop model in which the GT 
relation, $g_f/{\cal M}_f = 1/f_{f\!\bar f}$,
is necessary for reproducing the $\gamma\gamma$ 
anomaly amplitudes (\ref{ChLimAmp}) and (\ref{AnomAmpl}).
In the context of the coupled SD-BS approach, with its
dynamically dressed quarks and BS vertices, the GT relation for
quarks and Goldstone bosons is given by the chiral-limit
relation (\ref{ChLimSol}) (see, {\it e.g.}, \cite{Tandy97}).
The chiral-limit analytic derivation 
of Eq. (\ref{ChLimAmp}) from (\ref{ChLimSol}),
with its subtle interplay and cancellations between the bound-state vertex, 
WTI-preserving $qq\gamma$ vertices and dynamically dressed propagators,
transparently demonstrates the way the GT relation works for 
$\gamma\gamma$-decays in this context.
For massive pions, the $\gamma\gamma$-amplitude must be
evaluated numerically, but in fact changes very little,
implying that the GT relation continue to hold very accurately. 
As just argued above, in {\it ii)}, the GT relation should still hold 
as a rough approximation for the $s\bar s$ pseudoscalar bound state.
This, together with Eq. (\ref{ChLimAmp}), implies that
${\widetilde T}_{s\bar s} \sim N_c/(2\sqrt{2}\pi^2 f_{s\bar s})$.
For $f_{s\bar s} = 1.47\, f_\pi$ which we obtained in the model
\cite{jain93b}, this gives the GT relation-based estimate
${\widetilde T}_{s\bar s} \sim 0.68 {\widetilde T}_{u\bar u}$.
This is indeed in expected rough agreement with the accurate,
numerically obtained prediction of the model \cite{jain93b}, that
${\widetilde T}_{s\bar s}(0,0)=0.62\, {\widetilde T}_{u\bar u}(0,0)
=N_c/(2\sqrt{2}\pi^2 \, 1.61 f_\pi)$. This shows that our result is
quite reasonable. We remark that any model which is successful enough to
reproduce empirical values of $f_\pi$ and $f_{K^+}$, should
give a value for $f_{s\bar s}$ close to ours, since anything very
different from the estimate $f_{s\bar s}\sim f_\pi+2(f_{K^+}-f_\pi)$
would be unreasonable. Bound state descriptions that would be obtained
by using kernels supposedly better than ours (improved beyond the ladder
approximation by, say, including fully the {\it gluon} anomaly), 
must retain the good feature of agreeing approximately with the 
GT relation. This means that improving interaction kernels and, 
consequently, $f\bar f$ bound states, would not change
very much the $\gamma\gamma$-amplitudes with respect to our
${\widetilde T}_{f\!\bar f}$ even in the $s$-quark sector.

Regardless of any specific model realization of the coupled
SD-BS approach, Eqs. (\ref{f8}) and (\ref{f0}) with
${\widetilde T}_{s\bar s}(0,0) \leq {\widetilde T}_{\pi^0}(0,0)$ 
imply the following bounds on ${\bar f}_{\eta_8}$ and ${\bar f}_{\eta_0}$.
The equality holds when the chiral limit is applied to all
three flavors, implying that $f_\pi$
is the upper bound for ${\bar f}_{\eta_8}$ and lower bound for
${\bar f}_{\eta_0}$. As the $s$-quark mass
grows,  ${\widetilde T}_{s\bar s}(0,0)$  gradually diminishes,
so that the lower bound for ${\bar f}_{\eta_8}$ is $0.6 f_\pi$,
and the upper bound for ${\bar f}_{\eta_0}$ is $1.2 f_\pi$.

Let us now address the meaning of the apparent contradiction between 
the results of the coupled SD-BS approach on ${\bar{f}}_{\eta_8}/f_\pi$
and the corresponding results of $\chi$PT, as well as possible ways to 
-- at least in principle -- overcome it. To this end, let us recall 
Pham's paper \cite{Pham90} on $q\bar q$-loop corrections $\delta$ to 
the $\chi$PT result on ${\bar{f}}_{\eta_8}/f_\pi$. In the notation
of our Eq. (\ref{eta8Ampl}), he gets
\begin{equation}
 T_{\eta_8}(0,0) = \frac{f_\pi}{f_{{\eta_8}}}\, (1 - \delta) \,
\frac{T_{\pi^0}(0,0)}{\sqrt{3}} \, ,
\label{PhamCorr}
\end{equation}
where the axial-current decay constant $f_{{\eta_8}}$
appears. From the standpoint of our approach and its Eq.
(\ref{eta8Ampl}) relating the $\gamma\gamma$-amplitudes of $\eta_8$ 
and the chiral pion, $f_{{\eta_8}}/(1 - \delta)$ obviously
corresponds to {\it our} ${\bar{f}}_{\eta_8}$. Can $\delta$ resolve 
the discrepancy between our ${\bar{f}}_{\eta_8}$ and 
${\bar{f}}_{\eta_8}=f_{{\eta_8}}=1.25 f_\pi$ of $\chi$PT 
\cite{DonoghueHolsteinLin}?
Pham's rough estimate is $-0.28 < \delta < -0.19$. 
This would practically reduce the $\chi$PT
result from $1.25 f_\pi$ down to $f_\pi$.
In addition, his (following Ref. \cite{Kitazawa}) result
\begin{equation}
\delta = - \frac{8}{3}\, \left( 1 + \ln\frac{\Lambda^2}{M^2} \right)
                         \left( \frac{m_s \, M}{\Lambda^2} \right)
\label{delta}
\end{equation}
can be even more negative than $-0.28$, because there are large
uncertainties in the value of his cutoff $\Lambda$ (which can
be lower than the lowest value of 1.2 GeV used by \cite{Pham90}),
and in his effective $s$-quark mass $M$. Besides, the current
$s$-quark mass $m_s$ can be higher \cite{PDG94} than $m_s=175$ MeV
used by Pham. Each of these possibilities would make $\delta$ more negative.
On the other hand, the $\chi$PT results \cite{DonoghueHolsteinLin},
${\bar{f}}_{\eta_8}=f_{{\eta_8}}$ on the equality of the axial
and $\gamma\gamma$ decay constants, and
$f_{{\eta_8}} = (1+\alpha_8)f_\pi = 1.25 f_\pi$
on SU(3)$_f$ breaking effects in $f_{\eta_8}$,
are obtained in the one-loop approximation.
Higher loops can introduce significant changes,
including polynomial terms in meson masses, which
may be both large with respect to chiral logarithms,
and also not unambiguous \cite{DH89,DonoghueHolsteinLin}.
It is thus possible that $\alpha_8 < 0.25$ after all, which could make
even easier for the correction factor $(1-\delta)$ to reduce
${\bar{f}}_{\eta_8}$, as defined by us in (\ref{eta8Ampl}),
even below $f_\pi$.

Large uncertainties in the quantities entering in the estimate for
$q\bar q$ correction (\ref{delta}), obviously leave plenty of room
for usage of $q\bar q$ bound state models such as ours, which properly
embed the chiral behavior of the underlying theory. Including meson 
loops in $q\bar q$ bound state approaches is a difficult task and 
implies going beyond the ladder approximation, but 
it would help further diminish the gap that Pham \cite{Pham90}
started closing from the side of $\chi$PT by incorporating into
it the corrections due to quark loops. 

{\subsection{$\eta,\eta^\prime\to\gamma\gamma$ decay widths}}

The $\eta , \eta^\prime \to \gamma\gamma$ amplitudes are given in
terms of the $\gamma\gamma$-amplitudes of $\eta_8$ and $\eta_0$ as
	\begin{eqnarray}
	T_\eta(0,0)
	&=&
	\cos\theta\, T_{\eta_8}(0,0) - \sin\theta\, T_{\eta_0}(0,0)~,
	\\
	T_{\eta^\prime}(0,0)
	&=&
	\sin\theta\, T_{\eta_8}(0,0) + \cos\theta\, T_{\eta_0}(0,0)~.
	\end{eqnarray}
Expressing $T_{\eta_8}(0,0)$ and $T_{\eta_0}(0,0)$
through the $\gamma\gamma$-decay constants $\bar{f}_{\eta_8}$
(\ref{f8}) and $\bar{f}_{\eta_0}$ (\ref{f0}), we arrive at the 
standard ({\it e.g.}, see \cite{Donoghue+alKnjiga}) formulas for
the $\eta$ and $\eta^\prime$ decay widths:
	\begin{eqnarray}
	       W(\eta\to\gamma\gamma)
	&=&
	\frac{\alpha_{\rm em}^2}{64\pi^3}
	\frac{M_\eta^3}{3f_\pi^2}
	\left[
		\frac{f_\pi}{{\bar f}_{\eta_8}} \cos\theta
		-
		\sqrt{8} \frac{f_\pi}{\bar{f}_{\eta_0}} \sin\theta
	\right]^2~,
	\label{etawidth}
	\\
	       W(\eta^\prime\to\gamma\gamma)
	&=&
	\frac{\alpha_{\rm em}^2}{64\pi^3}
	\frac{M_{\eta^\prime}^3}{3f_\pi^2}
	\left[
		\frac{f_\pi}{{\bar f}_{\eta_8}} \sin\theta
		+
		\sqrt{8} \frac{f_\pi}{\bar{f}_{\eta_0}} \cos\theta
	\right]^2~.
	\label{etaprimewidth}
	\end{eqnarray}
The version of (\ref{etawidth}) and (\ref{etaprimewidth}) in which 
the axial-current decay constants $f_{\eta_8}$ and $f_{\eta_0}$
appear in place of ${\bar f}_{\eta_8}$ and ${\bar f}_{\eta_0}$,
requires a derivation where PCAC and soft meson technique is
applied to $\eta$ and $\eta^\prime$ \cite{FarrarGabag}.
These assumptions are impeccable for pions (leading to 
$f_\pi = {\bar f}_\pi$), but {\it not} for the 
$\eta$-$\eta^\prime$ complex. In fact, the latter is quite dubious 
for the heavy $\eta^\prime$ \cite{FarrarGabag}.
We do not need and do not use these assumptions since we 
{\it directly} calculate the ${\eta_8},{\eta_0}\to\gamma\gamma$ 
amplitudes, {\it i.e.},  ${\bar f}_{\eta_8}$ and ${\bar f}_{\eta_0}$. 
We also calculate $f_{\eta_8}$ and $f_{\eta_0}$  
independently of the $\gamma\gamma$-processes.

For the values of $\bar{f}_{\eta_8}$ and $\bar{f}_{\eta_0}$ obtained 
in our model choice \cite{jain93b}, namely (\ref{f8value}) 
and (\ref{f0value}), the best achievable consistency with the present
overall fit \cite{PDG96} to the experimental widths,
\begin{equation}
	 W^{\mbox{\scriptsize\rm exp}}(\eta\to\gamma\gamma)=
(0.46\pm 0.04)\keV \quad {\rm and} \quad
	 W^{\mbox{\scriptsize\rm exp}}(\eta^\prime\to\gamma\gamma)=
(4.26\pm 0.19)\keV,
\label{Wexp}
\end{equation}
then occurs for
$\theta \equiv \theta^{\mbox{\scriptsize\rm exp}}=-12.0^\circ$
(obtained through $\chi^2$ minimization). Then
	\begin{eqnarray}
	       W(\eta\to\gamma\gamma)
	&=&0.54~\keV~,
	\\
	       W(\eta^\prime\to\gamma\gamma)
	&=&5.0~\keV~.
	\end{eqnarray}
However, the present approach is capable of {\it predicting} 
the mixing angle $\theta$, and it remains to be seen if the predicted
$\theta$ can be close to the angle favored by the experimental
$\gamma\gamma$-widths. 

The issue of predicting $\theta$ will be addressed 
in the next section. There is another mixing-independent 
quantity related to ${\bar f}_{\eta_8}$ and ${\bar f}_{\eta_0}$, 
which we can predict before  predicting $\theta$.
It is the $R$-ratio, which is in fact measurable because
it is the combination of $\pi^0$, $\eta$ and $\eta^\prime$ widths:
\begin{equation}
R \equiv \left[\frac{
			W(\eta\to\gamma\gamma)}{M_\eta^3}
	+
	\frac{
		      W(\eta^\prime\to\gamma\gamma)}{M_{\eta^\prime}^3}
	\right]
	\frac{M_\pi^3}{
			       W(\pi^0\to\gamma\gamma)}
=\frac{1}{3}
	\left( \,
	      \frac{ f_{\pi}^2}{{\bar f}_{\eta_8}^2}
	    + 8 \, \frac{ f_{\pi}^2}{{\bar f}_{\eta_0}^2}
    \,  \right)~,
\label{Rratio}
\end{equation}
which is presently not known with satisfactory precision;
\cite{Anulli+al} quotes 
$R_{exp} = 2.5 \pm 0.5 (\mbox{\rm stat}) \pm 0.5 (\mbox{\rm syst})$.
A more precise value of the $R$-ratio (\ref{Rratio})
should come with DA$\Phi$NE's operation 
at its higher energy $\sqrt{s}=0.15~\GeV$, as this will enable
the measurement of $\gamma\gamma\to\eta^\prime$ \cite{Anulli+al}.

Since $R$ is independent of $\theta$, it will
most cleanly test our predictions (\ref{f8}), (\ref{f0}).
Precisely determined $R_{exp}$ can also help finding out
whether {\it (a)} ${\bar{f}}_{\eta_8} < f_\pi$, as follows 
in the coupled SD-BS approach from 
${\widetilde T}_{s\bar s}(0,0) < {\widetilde T}_{u\bar u}(0,0)$,
or {\it (b)} ${\bar f}_{\eta_8}= {f}_{\eta_8}\ge f_\pi$
as in $\chi$PT \cite{DonoghueHolsteinLin,Donoghue+alKnjiga}.
The present model gives us
${\widetilde T}_{s\bar s}(0,0) = 0.62\, {\widetilde T}_{u\bar u}(0,0)$,
so that $R = 2.87$ is obtained, which is well within the error
bars of the present experimental average \cite{Anulli+al}.
(Taking the chiral limit also for $s$-quarks would give
$R=3$, which is the upper bound for (\ref{Rratio}) in the present
approach. $R=3$ is still consistent with $R_{exp}$ \cite{Anulli+al}
within the present experimental accuracy.)

We should note: {\it i)} ${\widetilde T}_{s\bar s}$
is a quantity which can be especially practically used in
conjunction with a more accurate $R_{exp}$ to narrow down
the choice of models suitable for describing $\eta_8$ and $\eta_0$ 
(and ultimately $\eta$ and $\eta^\prime$). 
{\it ii)} Precise experimental determinations of $R$ can help
to find out if there are other admixtures $|X\rangle$
[{\it e.g.}, gluonium $|gg\rangle$,
$\eta(1295)$, $\eta(1440)$ (former $\iota$), $\eta_c$, ... ]
to $\eta$ and $\eta^\prime$, on top of the mixture
of $\eta_8$ (\ref{eta8def}) and  $\eta_0$ (\ref{eta0def}).
We can see both {\it i)} and {\it ii)} if we use our
predictions for ${\bar f}_{\eta_8}$ and ${\bar f}_{\eta_0}$,
(\ref{f8}) and (\ref{f0}), in (\ref{Rratio}), yielding
\begin{equation}
R = \frac{25}{9} + \frac{2}{9} \,
\left[\, \frac{2\sqrt{2}\pi^2 f_\pi}{N_c}\, {\widetilde T}_{s\bar s}(0,0)\,
\right]^2
\, ,
\label{ourR}
\end{equation}
which shows that in our approach $R$ depends only on one 
{\it variable{\footnote{As clarified above, getting
a reasonable $f_\pi$ first is obligatory for applications 
to $\gamma\gamma$-processes.}}}
model-dependent quantity: ${\widetilde T}_{s\bar s}$.
This is because
${\bar f}_{\eta_8}$ (\ref{f8}) and ${\bar f}_{\eta_0}$ (\ref{f0})
follow from the fact that, for {\it any} interaction kernel and
resulting propagator and bound state solutions,
${\widetilde T}_{\pi^0}(0,0) \equiv
{\widetilde T}_{u\bar u}(0,0)= N_c / (2\sqrt{2}\pi^2 f_\pi)$
in the chiral limit, and that this result remains an excellent
approximation for realistic $m_u$ and $m_d$ leading to empirical
$M_\pi$. Therefore, important variations in our predictions for
${\bar f}_{\eta_8}$ and ${\bar f}_{\eta_0}$, and thus $R$ (\ref{ourR}), 
can come only from ${\widetilde T}_{s\bar s}(0,0)$. The accuracy of 
${\widetilde T}_{s\bar s}(0,0)$ depends on the quality of the bound-state 
solution, but regardless of concrete model choices and results, 
the general inequality
${\widetilde T}_{u\bar u}(0,0)>{\widetilde T}_{s\bar s}(0,0)>0$ enables 
Eq. (\ref{ourR}) to provide the bounds $3 > R > 25/9 = 2.777...$ .
Hence, if experiments establish $R < 25/9$ by a significant amount, 
this will most probably indicate that in $\eta$ and $\eta^\prime$ there 
are admixtures ({\it e.g.}, glueballs) to $\eta_8$ (\ref{eta8def}) and
$\eta_0$ (\ref{eta0def}) which are ``inert" with respect to the 
interactions with photons, because this can lower the bound $R > 25/9$ 
most efficaciously. We will be able to address this in more detail after 
the discussion of the mixing, presented in the next section.

\section{Coping with mixing of etas in coupled SD-BS approach}
\label{ETA-ETAPRIME}

\noindent The mixing angle $\theta$ is often inferred from the 
empirical $\gamma\gamma$ decay widths of $\eta$ and $\eta^\prime$. 
This is how we established -- in Sec.~\ref{2GAMMA} --
that $\theta^{\mbox{\scriptsize\rm exp}}=-12.0^\circ$ 
is the empirically preferred mixing angle for the 
values of $\bar{f}_{\eta_8}$ and $\bar{f}_{\eta_0}$ obtained
in our model, namely (\ref{f8value}) and 
(\ref{f0value}). On the other hand, the angle $\theta$ is predicted  
by diagonalizing the $\eta$--$\eta^\prime$ mass matrix  evaluated 
in the $\eta_8$-$\eta_0$ basis -- such as 
the one predicted by our SD-BS approach and given in Eq. (\ref{MABJ0}) 
below. Obviously, for a satisfactory model description of the 
$\eta$--$\eta^\prime$ complex, the latter procedure should give the 
mixing angle close to the angle $\theta^{\mbox{\scriptsize\rm
exp}}$ required by the $\eta,\eta^\prime\to\gamma\gamma$ widths. 

In the $\eta$-$\eta^\prime$ complex, subtleties arise from the
interplay of the mixing due to the ${\rm SU(3)}_f$ breaking with the
$\mbox{\rm U}_{\rm A}{\rm (1)}$ {\it gluon} axial ABJ anomaly,
which couples to the flavor-singlet $\eta_0$ and removes the
nonet symmetry. (For a simple introduction, 
see Sec. 12.8 of \cite{Miransky} 
and Secs. III-3, VII-4 and X-3 of \cite{Donoghue+alKnjiga}).
Namely, in the coupled SD--BS approach,
where the states with good ${\rm SU(3)}_f$ quantum numbers
are constructed from the $f\bar f$ bound states $(f=u,d,s)$
obtained in Sec. \ref{QBARQ},
the eta (mass)$^2$ matrix $\hat{M}^2$ in the 
$\eta_8$--$\eta_0$ basis  
(\ref{eta8def})-(\ref{eta0def}) is given by
\begin{equation}
\hat{M}^2
=
\left[
	\begin{array}{cc}
	M_{88}^2 \,  & M_{80}^2 \\
	M_{08}^2 \,  & M_{00}^2
	\end{array}
\right]
=
\left[
	\begin{array}{cc}
	\frac{2}{3}(M_{s\bar{s}}^2 + \frac{1}{2} M_\pi^2)   \, \,
	&
     \,   \frac{\sqrt{2}}{3} (M_\pi^2 - M_{s\bar{s}}^2)
	\\
	\frac{\sqrt{2}}{3} (M_\pi^2 - M_{s\bar{s}}^2)       \, \,
	&
     \,   \frac{2}{3}(\frac{1}{2} M_{s\bar{s}}^2 + M_\pi^2)
	\end{array}
\right]
\label{MABJ0}
\end{equation}
{\it if} we neglect the gluon anomaly for the moment.
In agreement with other cases when the gluon ABJ anomaly is not
included, or is turned off, as in $N_c\to\infty$ limit
({\em e.g.}, \cite{Veneziano,Donoghue+alKnjiga}), the
diagonalization of (\ref{MABJ0}) yields an $\eta$ degenerate with
the pion, $M_\eta^2=M_\pi^2$, and without the $s\bar{s}$ component,
$\eta = \frac{1}{\sqrt 2}(u\bar{u} + d\bar{d})$,
whereas $\eta^\prime$ is a pure $s\bar s$ pseudoscalar, with
$M_{\eta^\prime}^2=M_{s\bar{s}}^2$. This happens at $\theta=-54.74^\circ$
and is obviously analogous to the ``ideal'' or ``extreme'' mixing which
is known to be a very good approximation for the mixing  of the vector
mesons $\omega$ and $\phi$. We can thus note in passing that the
present approach works well for the mixing of $\omega$ and $\phi$.
Nevertheless, this scenario is obviously catastrophic for the
$\eta$--$\eta^\prime$ system (the $\mbox{\rm U}_{\rm A}{\rm (1)}$
problem), so that gluon anomaly must be incorporated into our SD-BS
framework. Doing this on the fundamental level represents a
formidable task in any case, a task no-one has accomplished yet.
Moreover, an interaction kernel in the ladder approximation,
such as the simple gluon-exchange one that we have in the present model,
is inadequate for this task even in principle. Namely, by definition
it does not contain even the simplest annihilation graph
of a quark-antiquark pseudoscalar into two gluons
(and their recombination into another quark-antiquark pair) 
contributing to the processes such as the one in Fig. 2.
The contribution of the gluon ABJ anomaly operator
$\epsilon^{\alpha\beta\mu\nu} F^a_{\alpha\beta} F^a_{\mu\nu}$
to the $\eta_0$-mass, $M_{00}$, therefore cannot
be captured through a ladder kernel even in the roughest
approximation (leaving alone the issue of non-perturbative
gluon configurations such as instantons).

Therefore, some additional ingredients or assumptions
must be introduced into the present model in order to cope with
the $\eta$--$\eta^\prime$ system. Since going beyond ladder
approximation is not within the scope of the present work,
the following scheme is the most sensible at this level:
note that there is a standard way 
(see, {\it e.g.}, \cite{Miransky,Donoghue+alKnjiga})
to account for the anomaly effect
by {\it parametrizing} it through the term $\lambda_\eta$ added to the
$\eta_0$ mass, since only this singlet combination (\ref{eta0def}})
is coupled to the gluon anomaly, so that only its mass is affected by it. 
This corrects the $\mbox{\rm U}_{\rm A}{\rm (1)}$ problem
arising in the mass matrix evaluated with the nonet 
${\rm SU(3)}_f$--states $\eta_8$ and $\eta_0$. Let 
us do the same in our mass matrix (\ref{MABJ0}):
	\begin{equation}
	M_{00}^2
	\to
	\frac{2}{3} (\frac{1}{2} M_{s\bar{s}}^2 + M_\pi^2 )
	+
	\lambda_\eta~.
\label{Add}
	\end{equation}

Of course, parametrizing the effect of the gluon anomaly is far from 
actually calculating it unambiguously. In particular, the quantities 
we calculated for the $\eta_0$ under the assumption of nonet symmetry, 
$f_{\eta_0}$ and ${\bar f}_{\eta_0}$, are in fact also affected by the 
coupling of the gluon anomaly to $\eta_0$.
However, due to the large $N_c$ 
arguments, it makes sense to break nonet symmetry 
only on the level of the mass-shift parameter $\lambda_\eta$
while keeping our $\eta_0$ built of the same $f{\bar f}$
bound-state vertices as $\eta_8$, to which gluon anomaly
does not couple. This is because the gluon anomaly is
in the large $N_c$ limit suppressed \cite{Veneziano,Donoghue+alKnjiga}
as $1/N_c$, so that in our $f_{\eta_0}$ 
and ${\bar f}_{\eta_0}$ calculated
within the nonet scheme, only the contributions
of the order ${\cal O}(1/N_c)$ are missed.  
Our scheme is therefore a controlled approximation
on the level of large $N_c$ arguments.

The results obtained below for the mixing-dependent 
$\eta, \eta^\prime \to \gamma\gamma$ widths also turn out 
to be reasonable, providing an {\it a posteriori}
justification for our scheme. In the light 
of large $N_c$ arguments, such reasonable results are not
accidental and can be expected beforehand.

Let us also note that our assumptions are in
fact shared by many other approaches, explicitly
or implicitly. {\it E.g.}, Gilman and Kauffman \cite{GilKauf}
employ in their analysis nonet symmetry or broken
version thereof, pointing out that it is at least 
implicitly assumed by all who use the quark basis 
not differentiating between quark states belonging 
to the singlet from those belonging to the octet. 
Moreover, imposing the nonet breaking via introducing 
the additional parameter  $\lambda_\eta$
is basically the same way in which nonet symmetry is
broken in the chiral perturbation theory ($\chi$PT).  
In $\chi$PT, one faces the problem of how to
incorporate $\eta_0$, shifted upwards in mass by the gluon anomaly,
into the scheme that should involve Goldstone pseudoscalar mesons.
Bijnens, Bramon and Cornet \cite{BBC90} comment on the problems
encountered when working with this ninth state,
but stick to what they did earlier \cite{BBC88},
namely including $\eta_0$ ($\eta_1$ in their notation)
``in a simple nonet-symmetry context". Their Ref. \cite{BBC88}
conveniently parametrized the  nonet of (pseudoscalar)
Goldstone bosons in terms
of nine fields entering in the lowest order Lagrangian
consistent with current algebra and explicit breaking by the quark
masses, but the effect of the breaking of $U(1)_A$
is included only via {\it an extra mass term} for $\eta_0$. 
This is justified if one relies on large $N_c$ arguments, since 
$\eta_0$ is indeed a Goldstone boson in the limit $N_c\to\infty$ 
\cite{Ec89,Donoghue+alKnjiga}, and then $\eta_0$-mass is introduced 
as an extra parameter on top of that, which basically corresponds 
to our scheme.
 
Precisely in the light of $\chi$PT, our result (\ref{f0value})
for ${\bar f}_{\eta_0}$ appears very reasonable in 
spite of missing the contributions of ${\cal O}(1/N_c)$,
supporting our relying on the nonet symmetry scheme.
Namely, it is in excellent 
agreement with the results of $\chi$PT, being right between 
${\bar{f}}_{\eta_0}\approx 1.1 f_\pi$ quoted by \cite{BBC90}
and ${\bar{f}}_{\eta_0}= (1.04\pm 0.04) f_\pi$ of
\cite{DonoghueHolsteinLin}. 

Finally, the robustness of 
our $\gamma\gamma$-amplitudes to kernel variations,
resulting from the good chiral features of SD-BS approach
(as explained in Sec. \ref{2GAMMA}.B), 
also supports our scheme.

We therefore pursue the procedure of removing the 
$\mbox{\rm U}_{\rm A}{\rm (1)}$  problem by lumping the effects of
the gluon anomaly into a single $\eta_0$-mass shift parameter
$\lambda_\eta$ as in (\ref{Add}). Then, with our result
$M_{s\bar{s}}=0.721~\GeV$,
and with the experimental pion mass  $M_{\pi^0}=0.135$ GeV
(which the present SD-BS approach readily reproduces when
the strict chiral limit is relaxed \cite{jain93b}),
and with the choice $\lambda_\eta=1.165~\GeV^2$,
we get $\theta=-12.7^\circ$, which is 
very close to $\theta^{\mbox{\scriptsize\rm exp}}=-12.0^\circ$
favored by the empirical $\gamma\gamma$ widths.
Moreover, we then reproduce  the experimental value of $\eta$-mass,
$M_\eta = M_\eta^{exp} = 0.547$ GeV.
Admittedly, the $\eta^\prime$ mass is then somewhat too high,
$M_{\eta^\prime}=1.18~\GeV$. Of course, we could in principle 
pick such a value of $\lambda_\eta$, that this other mass 
$M_{\eta^\prime}$ would be reproduced, or still another
value ($\lambda_\eta=0.677~\GeV^2$) to reproduce empirical $M_\eta^2 +
M_{\eta^\prime}^2$. Naturally, $M_\eta$ would then be spoiled, 
but this is not the main reason why the two latter possibilities
are disfavored in the present approach. The main problem is that they
yield so negative values of the mixing angle ($\theta=-21.4^\circ$ when 
$M_\eta^2 + M_{\eta^\prime}^2$ is fitted, and $\theta=-22.8^\circ$ when 
$M_{\eta^\prime}^2$ is fitted to experiment), that they are incompatible 
with the present approach. In the present model \cite{jain93b}, so negative
mixing angles obviously yield unacceptable $\gamma\gamma$ widths, since  
the empirical $\gamma\gamma$ widths favor the mixing 
angle $\theta^{\mbox{\scriptsize\rm exp}}=-12.0^\circ$, and this speaks 
in favor of the first possibility, $\lambda_\eta=1.165~\GeV^2$, 
leading to $\theta = -12.7^\circ \approx \theta^{\rm exp}$.
Still, this $\theta^{\mbox{\scriptsize\rm exp}}=-12.0^\circ$
is the consequence of the particular model choice \cite{jain93b}
which led to the value (\ref{f8value}) and (\ref{f0value}) for
$\bar{f}_{\eta_8}$ and $\bar{f}_{\eta_0}$, respectively. Although 
we explained in the previous section why ${\widetilde T}_{s\bar s}$
(and consequently $\bar{f}_{\eta_8}$ and $\bar{f}_{\eta_0}$) must
be relatively stable to model kernel variations,
it is desirable to have a criterion which is even less
model-dependent. And indeed, we do have a reason why the coupled
SD-BS approach {\it in general} prefers the first procedure leading
to larger values of $\lambda_\eta$, and, consequently, less
negative values of $\theta$. Namely, it turns out that since
in the coupled SD-BS approach ${\bar f}_{\eta_8} < f_\pi$ for
any realistic value
of strange quark mass, the consistency with the experimental
$\eta,\eta^\prime\to\gamma\gamma$ widths  
is possible in this  approach
only for mixing angles less negative than roughly $- 15^\circ$.
This is easily seen, for example, in Fig.~1. of Ball {\it et
al.} \cite{Ball+al96}, where the  values of 
${\bar f}_{\eta_{8(0)}}/f_\pi$ consistent with experiment
are given as a function of the mixing angle $\theta$.
[It does not matter that they in fact plotted
$f_{\eta_{8(0)}}/f_\pi$ and not ${\bar f}_{\eta_{8(0)}}/f_\pi$. 
Namely, they used Eqs. (\ref{etawidth})-(\ref{etaprimewidth})
for comparison with the experimental $\gamma\gamma$-widths,
just with $f_{\eta_{8(0)}}/f_\pi$ instead of 
${\bar f}_{\eta_{8(0)}}/f_\pi$, so that the experimental constraints
displayed in their Fig. 1 apply to whatever ratios are used in these 
expressions.] On the other hand, the more negative values 
$\theta \lsim -20^\circ$ give good $\eta,\eta^\prime\to\gamma\gamma$
widths in conjunction with the ratio ${\bar f}_{\eta_8}/f_\pi = 1.25$
obtained by \cite{DonoghueHolsteinLin} in $\chi$PT. However, our 
approach belongs among constituent quark ones. In the next section 
we discuss why considerably less negative angles, $\theta \approx
-14^\circ\pm 2$ \cite{BS90}, are natural for constituent quark 
approaches in general.

The procedure leading to $\theta=-12.7^\circ$ is also corroborated 
by the results of some different approaches -- most notably, by
the results of the instanton liquid model, where one can actually
calculate the gluon anomaly mass shift instead of parametrizing it.
As Shuryak \cite{ShuryBookRev} pointed out, the instanton-induced
interaction leads simultaneously to both light pion and heavy
$\eta^\prime$ -- {\it i.e.}, the dynamics provided
by instantons can take care of the effects of the {\it gluon}
axial anomaly {\it and} provide the light pseudoscalars as the
Goldstone bosons of D$\chi$SB.
While the instanton-induced interaction may therefore be the main 
candidate which in the future one may try to include in the interaction 
kernel of the coupled SD-BS equations, the results of Alkofer {\it et al.}
\cite{ANVZ89} in the framework of the instanton liquid model have
already indicated that such an inclusion could easily lead to a 
{\it calculated} $\lambda_\eta$ similar to its
present {\it parametrized} value. Namely, Alkofer {\it et al.} \cite{ANVZ89}
find that due to instantons, the U(1)$_A$-anomalous contribution
$(2N_f/f) N/V$ must be added to the $\eta$--$\eta^\prime$ mass matrix.
This term, corresponding to our $\lambda_\eta$, also has the value very
close to our $\lambda_\eta=1.165$ GeV$^2$; it is equal to 
approximately 1.1 GeV$^2$
for their standard instanton density $N/V = 1$ fm$^{-4}$ and their
pseudoscalar decay constant $f=91$ MeV. (Number of flavors $N_f=3$.)
This gives them $\theta\approx - 11.5^\circ$,
$M_\eta\approx 0.527$ GeV and $M_{\eta^\prime}\approx 1.172$ GeV,
which is very similar to our results. 

{\subsection {Values of the mixing-dependent quantities}}

\noindent Once the mixing angle $\theta$ has been fixed, 
the predictions for the axial $\eta$ and $\eta^\prime$ decay 
constants are found from Eqs. (\ref{feta})--(\ref{fetaprime}).
$\theta = - 12.7^\circ$ implies  $f_\eta=112.6$ MeV and 
$f_{\eta^\prime}=117.1$ MeV.
This agrees almost perfectly with Scadron's \cite{Scadron84}
estimates $f_\eta \approx 1.22 f_\pi$ and
$f_{\eta^\prime} \approx 1.28 f_\pi$ obtained from the
GT relations at the quark level for the strange-to-nonstrange
constituent mass ratio ${\cal M}_s/{\cal M}_{ud}\approx 1.5$ 
(and for $\theta$ advocated by Scadron \cite{Scadron84}, which is, 
interestingly, the same as our favored $\theta = - 12.7^\circ$.)
However, these values are somewhat higher than the
experimental values $f_\eta^{exp}=94\pm 7$ MeV \cite{behrend91} 
or $79\pm 9$ MeV \cite{TPC90} and $f_{\eta^\prime}^{exp}=89\pm 5$ MeV 
\cite{behrend91} or $96\pm 8$ MeV \cite{TPC90},
deduced (under certain theoretical assumptions discussed below)
by CELLO \cite{behrend91} and TPC/$2\gamma$ \cite{TPC90}
collaborations from the $Q^2$-dependence of their measured 
$\eta(\eta^\prime)\gamma^\star\gamma$ transition form factors 
$T_{\eta(\eta^\prime)}(0,-Q^2)$ (in our notation),
where $k^{\prime 2} = - Q^2\neq 0$ is the momentum-squared 
of the spacelike off-shell photon
$\gamma^\star$. The same TPC/$2\gamma$ reference \cite{TPC90}
quotes also another pair of experimental values, 
$f_\eta^{(exp2)}=91\pm 6 \,\MeV$ and $f_{\eta^\prime}^{(exp2)}=78\pm 5$ MeV,
which were obtained from the experimental decay amplitudes into 
two on-shell photons under the assumption that one can write  
$T_{\eta(\eta^\prime)}(0,0) = 1/4\pi^2 f_{\eta(\eta^\prime)}$
by analogy with the axial anomaly result (\ref{AnomAmpl}) for the pion. 
However, because of the large $s$-quark mass, as well as the masses of
$\eta$ and $\eta^\prime$ which are, respectively, 4 and 7 times 
larger than the pion mass, this procedure can yield only a rough 
qualitative estimate.

On the other hand, our value of $f_{\eta}$ is much closer not only to 
Scadron's \cite{Scadron84} estimates and to the value 
$f_{\eta}=114\, \MeV$ of an approach \cite{Burden+Qian+al} somewhat
related to ours, but also to the model-independent result of $\chi$PT,
that $f_{\eta} = 1.02 f_{\pi} (f_{K}/f_{\pi})^{4/3}$ \cite{G+L}.
For the experimental ratio $f_{K}/f_{\pi} = 1.22\pm 0.01$, this gives 
$f_{\eta}= (1.3\pm 0.05) \, f_{\pi} = 120\pm 5 \, \MeV$ \cite{G+L},
for which both CELLO \cite{behrend91} and TPC/$2\gamma$ \cite{TPC90}
results are too low.

The experimental values $f_\eta^{exp}$ and $f_{\eta^\prime}^{exp}$
were extracted from the CELLO \cite{behrend91} and TPC/$2\gamma$
\cite{TPC90} data on the 
transition form factors $T_{\eta(\eta^\prime)}(0,-Q^2)$
assuming that the pole mass $\Lambda_{\eta(\eta^\prime)}$
parametrizing their fit to the data, can be identified with 
$2 \pi \sqrt{2} f_{\eta(\eta^\prime)}$.
Then, the pole fits to the data could join smoothly (as $Q^2\to\infty$)
the perturbative QCD prediction \cite{BrodskyLepage} for 
$T_{\eta(\eta^\prime)}(0,-Q^2)$, {\it i.e.}, the pole fits would then 
agree not only with the QCD asymptotic form $1/Q^2$, but also with
its coefficient.
However, note that the values of $f_\eta^{exp}$ and
$f_{\eta^\prime}^{exp}$ quoted above, are {\it all}
close to $m_\rho/(2 \pi \sqrt{2})=86.4$ MeV, indicating that a 
connection with the vector-meson dominance interpretation 
(that $\Lambda_{\eta(\eta^\prime)}\approx m_\rho$) \cite{behrend91,TPC90} 
may indeed exist at the investigated
range of $Q^2$. On the other hand, since Gasser and Leutwyler's
model-independent calculation \cite{G+L}, Scadron's \cite{Scadron84}
GT estimates, Burden {\it et al.} \cite{Burden+Qian+al}, and the present 
approach, all agree that $f_{\eta(\eta^\prime)}$ should be noticeably 
larger than $f_\pi$, the extraction 
of $f_{\eta}^{exp}$ and $f_{\eta^\prime}^{exp}$ from the transition
form factors $T_{\eta(\eta^\prime)}(0,-Q^2)$ probably cannot be done 
accurately at the ranges of $Q^2$ investigated so far. 
That this is indeed so, is indicated by the experimental value \cite{PDG96}
$f_{\pi^0}=84.1\pm 2.8 \MeV$ ($\approx m_\rho/(2 \pi \sqrt{2})$ again), 
extracted by the same method. The central value is 10\% below well 
established $f_\pi^{exp}=92.4\pm 0.3$ MeV. This cannot be explained
by the small isospin violation, indicating that $f_\eta^{exp}$ and 
$f_{\eta^\prime}^{exp}$ could have been underestimated too. 

Our predictions
for the $\eta$ and $\eta^\prime$ two--photon widths are also totally
fixed now, being given by our ${\bar f}_{\eta_{8}}$ and
${\bar f}_{\eta_{0}}$ used in (\ref{etawidth}) and
(\ref{etaprimewidth}),
without any additional parameters to adjust.
Our preferred angle $\theta=-12.7^\circ$
leads to the predictions (displayed also in Table I.)
	\begin{eqnarray}
	       W(\eta\to\gamma\gamma)
	&=&
        0.561~\keV \, ,
\label{etaPredict}
	\\
	       W(\eta^\prime\to\gamma\gamma)
	&=&
	4.913~\keV~.
\label{etaPrimePredict}
	\end{eqnarray}
These predictions are at first sight not very successful 
since, according to Table I, our best predictions overshoot the present
\cite{PDG96} experimental averages (\ref{Wexp}) for
$\eta , \eta^\prime \to \gamma\gamma$ by some $20$\%.
However, we should not be dissatisfied with
these results because of the following:

\noindent {\it a)} Ball {\it et al.} \cite{Ball+al96} and,
in effect, Review of Particle Properties itself \cite{PDG96}
(referring to the note on p. 1451 of \cite{PDG94}), suggest that 
only the more recent data on $\eta, \eta^\prime \to \gamma\gamma$
should be retained, whereby the presently ``official" values
(\ref{Wexp}) are modified to \cite{PDG94,Ball+al96}
	\begin{eqnarray}
		 W^{exp}_{NEW}(\eta\to\gamma\gamma)
	&=&(0.510 \pm 0.026)~\keV~,
\label{WetaNEW}
	\\
		W^{exp}_{NEW}(\eta^\prime\to\gamma\gamma)
	&=&(4.53 \pm 0.59)~\keV~,
\label{WetaPrimeNEW}
	\end{eqnarray}
and these experimental values agree much better
with our predictions.

\noindent {\it b)} We did not vary any model parameters, but used 
the parameters obtained from Ref.'s \cite{jain93b}
broad fit to the meson spectrum and pseudoscalar decay constants.
This fit did not include $\eta$--$\eta^\prime$ system in any way, 
so that everything we calculated for it are pure predictions.

{\subsection{A side issue: speculations about other admixtures}}

\noindent In the present approach, $\eta_0$ and $\eta_8$ (and 
consequently $\eta$ and $\eta^\prime$) are constructed exclusively of 
the ground state pseudoscalar $q\bar q$ bound states. Nevertheless,
it has been often speculated about additional admixtures, notably 
glueballs. Farrar \cite{Farrar} points out that experiments 
appear to indicate that there is a glueball-like pseudoscalar which is
much lighter than estimated by quenched lattice calculations,
thus motivating us to
speculate on the consequences of such admixtures. 
We do {\it not} have at this point the ambition
to include such additional admixtures in our approach. However,
we can anyway look into some of the consequences
that such admixtures would have, by simply {\it assuming} that 
they were present besides the quarkonium $\eta_0$ and $\eta_8$ 
as constructed in this paper. 

Take for example the simplest and most usual assumption
\cite{Donoghue+alKnjiga}, that only the SU(3)$_f$ - singlet
(\ref{eta0def}) can be significantly modified in this way:
\begin{equation}
|\eta_0\rangle    \to 
 \frac{\cos\varphi}{\sqrt{3}} (|u\bar{u}\rangle + |d\bar{d}\rangle
					     + |s\bar{s}\rangle)
 + \sin\varphi |X\rangle~,
 \label{mixX}
\end{equation}
where $\varphi$ is the new mixing angle, 
a new parameter expressing the assumed strength of the 
unspecified admixture $|X\rangle$ into $\eta_0$.

If $|X\rangle$ is a state that does not couple to photons directly
({\it e.g.}, gluonium $|gg\rangle$),
the results for $\gamma\gamma$-decays will be modified in a
particularly simple way: in formulas (\ref{eta0Ampl}),
(\ref{etawidth}), (\ref{etaprimewidth}) and (\ref{Rratio}), one
should just replace $1/{\bar f}_0$ by $\cos\varphi/{\bar f}_0$.
This can reduce $R$ (\ref{ourR}) strongly, as
the largest term in Eq. (\ref{ourR}), 25/9, would then be
modified to $25/27 \, + \, \cos^2\varphi \, 50/27$.

We should also note that such an admixture (\ref{mixX}) would
help to fit the masses of both $\eta$ and $\eta^\prime$
to their experimental values precisely -- thanks to the 
new free parameter $\varphi$, of course. Eq. (\ref{mixX})  
modifies elements of the mass matrix to
\begin{equation}
M_{80}^2 \to
\cos\varphi \frac{\sqrt{2}}{3} (M_\pi^2 - M_{s\bar{s}}^2) \, ,
\label{modifM80}
\end{equation}
\begin{equation}
M_{00}^2 \to 
\cos^2\varphi \, \frac{2}{3}(\frac{1}{2} M_{s\bar{s}}^2 + M_\pi^2)
 + {\widetilde \lambda}_\eta  \,  ,
\label{modifM00}
\end{equation}
where  ${\widetilde \lambda}_\eta \equiv
		  \lambda_\eta + \sin^2\varphi \, M_X^2$
takes the place that $\lambda_\eta$ has for $\varphi = 0$.
If $|X\rangle$ is not a single state, but a mixture 
of various states, its mass $M_X$ has a meaning of an 
effective mass. 

The experimental masses $M_\eta = 547$ MeV and
$M_{\eta^\prime} = 958$ MeV, as well as the $\eta - \eta^\prime$
mixing angle $\theta = - 17.1^0$, are then obtained for
$\varphi = 42.43^0$ and ${\widetilde \lambda}_\eta =
(0.873 \, {\rm GeV})^2$. Nevertheless, it turns out that 
the fit to the data is still not improved as much as one would 
expect when an additional free parameter is introduced, so that we
did not detect indications for the need for an admixture of such 
states to what we have in the present model.
For example, our $R$-ratio then drops to $R=1.80$. 
This is much further from the present central
experimental value than $R$ predicted by our approach
without glueballs, but just in case that data from future precision
measurements may strongly violate our bound on the $R$-ratio, 
it is important to point out that -- at least from the standpoint
of our approach -- such a violation would be a strong indication
of the presence of some ``inert" admixture, like gluonium. 
At present, however, the data are consistent with the bound
$R > 25/9$ following generally from the SD--BS approach without
gluonium admixture, and even favor the value $R=2.87$ following
from the present concrete model choice \cite{jain93b} without
glueballs, over the value with the admixture quoted above.
Moreover, the $\eta\to\gamma\gamma$ width with the gluonium admixture 
improves only marginally, by 4\%, while the $\eta^\prime\to\gamma\gamma$
width gets {\it spoiled} by more than a factor of 2.

We therefore conclude that we found no indication that 
admixtures of glueballs, or other states with similar 
effects on $\gamma\gamma$-decays, would be favored by 
the present experimental data. Consequently, there is no 
strong motivation
for enlarging the present framework by finding solutions
for pseudoscalar glueballs and treating them on the same
footing as our pseudoscalar $q\bar q$ bound states. 
(It is amusing that 
 $\varphi=42.43^\circ$ in conjunction with the vanishing
gluon anomaly contribution, $\lambda_\eta = 0$, implies
$M_X = 1.294$ GeV -- practically the same as 
the mass of $\eta(1295)$. However, 
this can only be viewed as accidental at this point.)

\section{Summary, discussion and conclusions}

\noindent
The relativistically covariant constituent $q\bar q$ bound-state 
model \cite{jain93b} used here is consistent with current algebra 
because it incorporates the correct chiral symmetry behavior thanks
to D$\chi$SB obtained in an, essentially, Nambu--Jona-Lasinio (NJL)
fashion, but the model interaction is less schematic. 
Notably, when care is taken to preserve WTI of QED, it 
reproduces (in the chiral limit even analytically and independently
of the internal meson structure) the Abelian axial
anomaly results, which are otherwise notoriously difficult to 
reproduce in bound-state approaches (as illustrated by, {\it e.g.},
\cite{HorbKoniuk93} and especially references therein). Observables 
such as meson masses, $f_\pi, f_K, f_\eta, f_{\eta^\prime}$ and 
$\gamma\gamma$-decay amplitudes can be calculated without additional 
parameters after an {\it Ansatz} has been made for the gluon
propagator entering in the SD-BS equations, which are consistently 
coupled in the generalized (or improved) rainbow-ladder approximation
(in the terminology of, {\it e.g.}, \cite{Miransky} or \cite{Roberts}). 
However, to avoid U(1)$_A$-problem in the $\eta$-$\eta^\prime$
complex, we have to introduce an additional parameter, $\lambda_\eta$, 
representing the contribution of the gluon axial anomaly to the
mass of $\eta_0$, in analogy with the similar $\eta_0$-mass parameter 
in the $\chi$PT Lagrangian in Ref. \cite{BBC88}, for example. Since
the gluon anomaly contribution vanishes in the large $N_c$ limit
as $1/N_c$, our $q\bar q$ bound-state pseudoscalar mesons behave  
in the $N_c\to\infty$ and chiral limits in the same way  as those
in $\chi$PT ({\it e.g.}, see \cite{G+L} or \cite{Donoghue+alKnjiga}): 
as the strict chiral limit
is approached for all three flavors, the SU(3)$_f$ octet pseudoscalars 
{\it including} $\eta$ become massless Goldstone bosons, whereas the 
$\eta^\prime$-mass is of order $1/N_c$ since it is purely due to the 
gluon anomaly. In the $N_c\to\infty$ limit with nonvanishing quark 
masses, the ``ideal" mixing takes place so that $\eta$ consists of
$u,d$ quarks only and becomes degenerate with $\pi$, whereas $\eta^\prime$
is the pure $s\bar s$ pseudoscalar.

In our bound-state approach, $f_\pi$, ${\bar{f}}_{\eta_8}$, 
${\bar{f}}_{\eta_0}$, as well as ${f}_{\eta_8}$, ${f}_{\eta_0}$ 
and $f_\eta$ and $f_{\eta^\prime}$, 
are {\it all} calculated quantities,
while most other theoretical frameworks treat at
least one of them, $\bar{f}_{\eta_0}$, as a free parameter
(fixed together with $\theta$ from the experimental
widths of $\eta,\eta^\prime\to\gamma\gamma$).

Our prediction $f_{\eta_8}/f_\pi=1.31$ agrees rather well with 
$f_{\eta_8}/f_\pi=1.25$ of $\chi$PT \cite{DonoghueHolsteinLin}.
Nevertheless, this one-loop $\chi$PT calculation also lead to
the identification of their axial-current and $\gamma\gamma$-decay
constants, $f_{\eta_8}={\bar f}_{\eta_8}$, which differs from
our results on $\eta_8$.
More precisely, the observation that for realistic $s$-quark masses,
${\widetilde T}_{s\bar s}(0,0) < {\widetilde T}_{\pi^0}(0,0)$
always holds in the coupled SD-BS approach, leads to 
$\frac{3}{5}\, f_\pi <{\bar{f}}_{\eta_8} <f_\pi$
and $f_\pi <{\bar{f}}_{\eta_0} <\frac{6}{5}\, f_\pi$. 
These inequalities hold irrespective of the model
parameters or the quality of the interaction kernel.
${\bar{f}}_{\eta_8} = f_\pi  = {\bar{f}}_{\eta_0}$ is realized
in the chiral limit, whereas the opposite bounds are approached 
when the $s$-quark mass grows huge, leading to the decrease of
${\widetilde T}_{s\bar s}(0,0)/{\widetilde T}_{\pi^0}(0,0)$.
There is no disagreement with $\chi$PT
regarding ${\bar{f}}_{\eta_0}$, either concerning the general
bound $f_\pi <{\bar{f}}_{\eta_0} <\frac{6}{5}\, f_\pi$ of the
coupled SD-BS approach, or our result obtained using the concrete
model of \cite{jain93b}, namely ${\bar f}_{\eta_0}/f_\pi=1.067$.
This agrees well with the values found in $\chi$PT
\cite{DonoghueHolsteinLin,BBC90}.
The apparent contradiction between the results of the coupled SD-BS 
approach on $f_\pi/{\bar{f}}_{\eta_8}$, and the corresponding results 
of $\chi$PT was discussed in detail in Sec. \ref{2GAMMA}.B. 
Let us now address the intimately related issue of different preferred 
mixing angles in these respective approaches. 

In conjunction with the updated experimental widths 
(\ref{WetaNEW})-(\ref{WetaPrimeNEW}), ${\bar{f}}_{\eta_8} < f_\pi$
implies that the coupled SD-BS approach is compatible with
the mixing angles which are less negative than
$\theta\approx - 15^\circ$. For our concrete model choice \cite{jain93b}
and the resulting values (\ref{f8value})-(\ref{f0value})
of ${\bar{f}}_{\eta_8}$ and ${\bar{f}}_{\eta_0}$, the
favored value of $\theta$ is between the values accepted
till mid-eighties, namely $\theta\approx - 10^\circ$
determined from the SU(3)$_f$ breaking given by the
Gell-Mann--Okubo mass formula, and the lowest of the values
$\theta\in [-17^\circ, -20^\circ]$ favored nowadays 
\cite{Ball+al96,GilKauf,DonoghueHolsteinLin,G+L}.

In order to see that the mixing angles considerably less negative
than those in $\chi$PT ($\theta\sim -20^\circ$) are a natural and 
expected prediction in a constituent approach  such as ours,
it is instructive to recall the paper of Bramon and
Scadron \cite{BS90} where the mixing angle of 
$\theta = -14^\circ \pm 2^\circ$
follows from a rather exhaustive set of data if the SU(3)$_f$
breaking is taken into account in terms of the {\it constituent} quark
mass ratio ${\cal M}_s/{\cal M}_{ud} \approx 1.4 - 1.5$. 
SU(3)$_f$-breaking ratios
somewhere  around this interval are considered realistic because
they lead to good descriptions of many hadronic properties in numerous 
dynamical models; notably, close to this interval is also the ratio
($\approx 1.63$) of the constituent masses  $B_f(0)/A_f(0)$ generated by
D$\chi$SB in Jain and Munczek's approach. Bramon and Scadron \cite{BS90}
extracted   their average $\theta=-14^\circ \pm 2^\circ$
from the strong interaction tensor $T\to PP$ decays, and the vector
$V\to\gamma P$ and pseudoscalar $P\to\gamma\gamma$ radiative decays.
(And when extracted just from $\eta,\eta'\to \gamma\gamma$ pertinent 
here, and other SU(3)$_f$-breaking-ratio-dependent {\it radiative} 
decays, their angle is even lower, $-11^\circ \pm 2.4^\circ$.)
They point out that more negative values $\theta\sim - 20^\circ$ 
in the $\chi$PT framework are due to the 
way of implementing the SU(3)$_f$-breaking  
(through the values of the
decay constants $f_{\eta_8}$ and $f_{\eta_0}$), 
differing from that in the constituent-quark approaches.

Now, our SU(3)$_f$-breaking is fixed by Jain and Munczek's
\cite{jain93b}
choice of parameters, so that our calculated value of $\theta$
varies only if we vary $\lambda_\eta$ which parametrizes the
effects of the gluon anomaly.
In the light of Bramon and Scadron's \cite{BS90} observations discussed 
above, and the fact that that our SU(3)$_f$-breaking leads to the ratio
of strange-to-nonstrange constituent masses of 1.63,
it is understandable and expected that our constituent approach should
give reasonably good description of $\eta,\eta^\prime\to\gamma\gamma$
for angles less negative than in $\chi$PT; {\it i.e.}, it is no
longer surprising that our preferred angle turned out to be 
$\theta = - 12.7^\circ$. However, it is not only that these values 
are the preferred ones in our presently chosen model \cite{jain93b} 
because they are more empirically successful than other values.
In addition to that, since in the coupled SD-BS approach 
${\bar f}_{\eta_8} < f_\pi$ rather generally (for any realistic 
value of strange quark mass), the consistency with the experimental
$\eta,\eta^\prime\to\gamma\gamma$ widths is possible -- in our 
approach -- only for mixing angles less negative than roughly 
$- 15^\circ$, as already pointed out above.

That all this is in qualitative agreement with what was known from 
relatively simple-minded constituent-quark models even before the 
analysis of \cite{BS90}, can be seen, {\it e.g.}, from Zieli\' nski's 
review \cite{Z87} on radiative decays of mesons. He observed that in the 
scenarios that related apparent suppressions of radiative decays of strange
mesons to a larger mass of the $s$-quark, a significant suppression
of the annihilation amplitude of $s\bar s$ pairs into two
photons was also expected, and with the latter suppression of
order 0.5 relative to annihilation amplitudes of non-strange
quarks (relevant, {\it e.g.}, to the model of \cite{GodfIsgur85}),
the two-photon widths of both $\eta$ and $\eta^\prime$ could (however 
roughly) best be described with $\theta\sim -11^\circ$.
Remembering the limitations on mutually consistent $\theta$
and ${\bar f}_{\eta_{8(0)}}$, we see that our values of
$\theta$ and ${\bar f}_{\eta_{8(0)}}$ fit with our third element,
$ {\widetilde T}_{s\bar s}(0,0)= 0.62{\widetilde T}_{u\bar u}(0,0)$,
into a logical scheme which is consistent with
the behavior of the approaches similar to ours. 
Zieli\' nski \cite{Z87} also discussed how $\theta $ was much more
negative ($\sim -20^\circ$)  in  chiral theories,
but pointed out that the determination of the
pseudoscalar mixing angle was model dependent, and a clean-cut
choice among various schemes was rather difficult to establish.
Our discussion, and the results of, {\it e.g.}, 
Bramon and Scadron \cite{BS90}, Pham \cite{Pham90}
and Ball {\it et al.} \cite{Ball+al96}, shows that 
this assessment still holds, but also that there has
been some progress in narrowing the interval of possible mixing angles. 
In this connection, recall the observation of \cite{Ball+al96},
that newer experimental input [our Eqs. (\ref{WetaNEW})
and (\ref{WetaPrimeNEW})] reduces the mixing angle even more
than Pham \cite{Pham90} realized, to $\theta=-(17\pm 2)^\circ$.
This is not any more so far away from our preferred $\theta$
(especially considering that the value of the correction $\delta$,
Eq. (\ref{delta}), can be even more negative than Pham's values
\cite{Pham90}).
If various approaches succeed in including physical mechanisms they
have been missing so far, their predictions for $\theta$ will probably
tend to a unique value.
Likewise will be with ${\bar f}_{\eta_8}$ and ${\bar f}_{\eta_0}$.
In view of \cite{BS90,Pham90,Ball+al96},
this final value at which $\theta$ will settle, may well be roughly
in between the values favored nowadays by $\chi$PT and by quark
model approaches such as ours. Thus, $\theta \sim -14^\circ$
to $- 17^\circ$ maybe encompasses the final result. 
An ${\bar f}_{\eta_8}$ which would be rather close to the chiral limit 
value ${\bar f}_{\eta_8}=f_\pi$, because the chiral-loop contributions 
would be -- like in \cite{Pham90} -- to some extent (over)canceled by
some other contributions (like our bound-state strange mass-breaking
effects), would agree better with such a $\theta\sim  -14^\circ$ to 
$-17^\circ$. In our approach, the physical mechanisms which are now 
absent, are those corresponding to loops in $\chi$PT. Including them
obviously implies substantial enlargements beyond the present framework. 
However, this also holds for others -- {\it e.g.}, in $\chi$PT
one might pose the question what would the effects of higher loops and
vector mesons be. At present, no approach can claim to have all the 
relevant physics included, and therefore the ultimate values for 
$\theta$ and ${\bar f}_{\eta_8}$.

The present experimental value of the $R$-ratio (\ref{Rratio}) is 
described reasonably well by our approach. What if more 
precise measurements ({\it e.g.}, at DA$\Phi$NE \cite{Anulli+al})
constrain $R_{exp}$ below 25/9? A strong violation of this bound
(say, $R_{exp} < 2.5$) would indicate that important
admixtures other than $\eta_8$ and $\eta_0$ are present in 
$\eta$ and $\eta^\prime$. If the violation is not
that strong, the following possibility is also viable:
some of the values in the interval $2.5 < R_{exp} < 25/9$
can be satisfied by ${\bar{f}}_{\eta_8}$ and
${\bar{f}}_{\eta_0}$ predicted by $\chi$PT.
Hence, such smaller violation of our bound can also 
mean that the prediction of $\chi$PT,
that $f_\pi < {\bar{f}}_{\eta_8}$,
is favored over our prediction. This would indicate 
that in the case of the $\eta$--$\eta^\prime$ complex,
the ladder-approximated SD-BS approach makes larger error
by neglecting meson loops than, {\it e.g.},
in the case of the charge pion form factor calculated in 
in the context of SD equations,
where the contribution of meson loops was estimated
to be much smaller than that of the quark core \cite{Alkofer+al95}.

The quantities dependent on the $\eta$-$\eta^\prime$ mixing, namely 
axial-current decay constants $f_{\eta}$ and $f_{\eta^\prime}$, 
masses and $\gamma\gamma$-decay widths of $\eta$-$\eta^\prime$, 
are satisfactorily close to
data (or other theoretical predictions such as $\chi$PT)
considering that -- except for parametrizing the 
mass shift due to coupling of $\eta_0$ to nonabelian axial
anomaly -- we did not do any parameter fitting, but used the
parameters obtained from Jain and Munczek's \cite{jain93b}
broad fit to the meson spectrum and decay constants. We conclude 
that their model \cite{jain93b} again performed well.

Since the coupled SD-BS approach is -- due to the key role of 
D$\chi$SB -- akin to the NJL model conceptually, the progress we 
made is best illustrated through the comparison with the analysis 
of the $\pi^0, \eta \to \gamma\gamma$ decays and properties of the 
pion, kaon and $\eta$, performed in a NJL model (extended to include 
three flavors and the 't Hooft determinantal instanton-induced 
interaction) in \cite{TO95} and in parts of \cite{NTO96,NTO97}.
($\eta^\prime$ was not treated in \cite{TO95,NTO96,NTO97}.)

For the choice of model parameters preferred by Takizawa, Oka, and Nemoto 
\cite{TO95,NTO96,NTO97}, the experimental amplitude for $\eta \to \gamma
\gamma$ is reproduced, but the $\eta$-mass is 7\% below the experimental 
value. The mixing angle is $\theta = -1.25^\circ$, showing that their 
$U(1)_A$-breaking is stronger than in our approach (not to mention 
the one in $\chi$PT), forcing their $\eta$ to be an almost pure $\eta_8$.
Their kaon decay constant $f_K = 96.6$ MeV is 15\% below the observed one. 
Accordingly, the predicted $\eta$ decay constant, $f_\eta \approx f_\pi$, 
is uncomfortably far from what the model-independent result of $\chi$PT 
\cite{G+L}, $f_{\eta} = 1.02 f_{\pi} (f_{K}/f_{\pi})^{4/3}$, gives when 
the empirical $f_K/f_{\pi}$ is plugged in. 

While our results compare rather favorably with the above, 
the best examples of advantages both 
in the conceptual consistency and in the quantitative details 
which coupled SD-BS approach has with respect to the 
NJL-model, are $P \to\gamma\gamma$ decays.
Namely, we must criticize the seemingly successful 
reproduction of the anomalous $\pi^0,\eta\to\gamma\gamma$ 
amplitudes by \cite{TO95,NTO96,NTO97}. 
In contradistinction to the coupled SD-BS approach
with nonlocal interactions -- in particular Jain and Munczek's 
model, where the UV cutoff is either not needed (in the chiral 
limit \cite{jain91}) or practically infinite compared 
to the relevant hadronic scales -- the NJL approach contains
a low cutoff. In spite of this, Refs. \cite{TO95,NTO96,NTO97}
leave the convergent integrals unregulated, because the triangle 
diagram reproduces the anomalous $\pi^0\to\gamma\gamma$ amplitude 
(\ref{AnomAmpl}) only if there is no NJL cutoff 
\cite{TO95,AlkoferReinhardtKnjiga,LKR96,PlantBirse97} -- 
else, an underestimate of, typically, 20\% occurs \cite{LKR96}.
While Refs. \cite{TO95,NTO97} claim the improvement of the 
$\eta \to \gamma\gamma$ amplitude and width with respect to the 
earlier treatment of Bernard {\it et al.} \cite{Bernard+al93},
the consistent viewpoint is that of \cite{Bernard+al93}: 
once the cutoff is introduced, the effective theory is
defined and should not be altered for the purpose of calculating
various quantities. In such effective theories, the missing part 
of the anomalous amplitude -- lost due to the cutoff -- should be 
found in additional diagrams \cite{AlkoferReinhardtKnjiga} which 
contribute since the cutoff cannot be let to infinity. {\it I.e.}, in 
the class of models employing only local interactions and therefore 
needing a low cutoff, a {\it simple} incorporation of the anomaly is 
not possible \cite{AlkoferReinhardtKnjiga,LKR96,PlantBirse97}, in 
contrast to the coupled SD-BS approaches employing also nonlocal 
interactions and thus not having such a cutoff.

Ref. \cite{Burden+Qian+al} is another approach to $q\bar q$ substructure 
incorporating D$\chi$SB, and it is even closer to us than the NJL model. 
The interaction used in \cite{Burden+Qian+al} is nonlocal, like ours, 
allowing the generation of momentum-dependent dynamical mass and BS-vertices,
so that there are no problems with a low cutoff like in the NJL model. 
The mixing angle they favor, $\theta\sim +5$, results from its
treatment as an external parameter on which the mass and other properties
of $\eta$ depend. However, their axial current decay constant 
$f_\eta=114$ MeV is close to ours.

Extending the treatment of the $\pi^0\gamma^\star\to\gamma$ transition
form factor of \cite{KeKl1,Lesna} to the
$\eta(\eta^{\prime}) \gamma^\star\to\gamma$ transition form factors
is presently under investigation \cite{KeKlProgress}.


\section*{Acknowledgments}
\noindent The authors acknowledge useful discussions with P. Jain
and the support of the Croatian Ministry of Science and
Technology contracts 1--19--222 and 009802.


\begin{table}
\begin{tabular}{cccc}
$P$ &  $
		  W(P\to\gamma\gamma)$ &
$
	     W^{exp}(P\to\gamma\gamma)$ & 
$ W^{exp}_{NEW}(P\to\gamma\gamma)$\\
\hline
$\pi^0$  & 7.7 & 7.74$\pm 0.56$
&$ not \, \, applicable $
\\
$\eta$          & $0.56\times 10^{+3}$  & $(0.46\pm 0.04)\times 10^{+3}$ & 
$(0.51 \pm 0.026)\times 10^{+3}$
\\
$\eta^\prime$   & $4.9\times 10^{+3}$   & $(4.26\pm 0.19)\times 10^{+3}$ & 
$(4.53 \pm 0.59)\times 10^{+3}$
\\
\end{tabular}
\caption{Comparison of the calculated $\gamma\gamma$ decay widths
(in eV) of 
$\pi^0, \eta$ and $\eta^\prime$
with their average experimental widths, as well as the experimental widths
$W^{exp}_{NEW}$ obtained when only more recent measurements are taken into
account.
The widths are calculated using the empirical masses 
in the phase-space factors in conjunction with calculated amplitudes.
The tabulated $\eta$ and $\eta^\prime$ calculated widths correspond to 
the case when their mixing adjusts the mass of $\eta$ to its 
empirical mass.}
\label{tab:resullts_01}
\end{table}


\newpage

\section*{Figure captions}

\begin{itemize}

\item[{\bf Fig.~1:}] The diagram for $P \to\gamma\gamma$
		decays ($P = \pi^0,\eta, \eta^\prime, ... $).
		Within the scheme of generalized impulse approximation,
		the propagators and vertices are dressed.

 \item[{\bf Fig.~2:}] The annihilation of the $f\bar f$ bound-state 
	vertex $\Gamma_{f\bar f}$ into two gluons and their recombination 
	      into the quark-antiquark bound state consisting of 
	      possibly different flavors $f^\prime {\bar f}^\prime$.

\end{itemize}


\newpage

\vspace*{4cm}
\epsfxsize = 18 cm \epsfbox{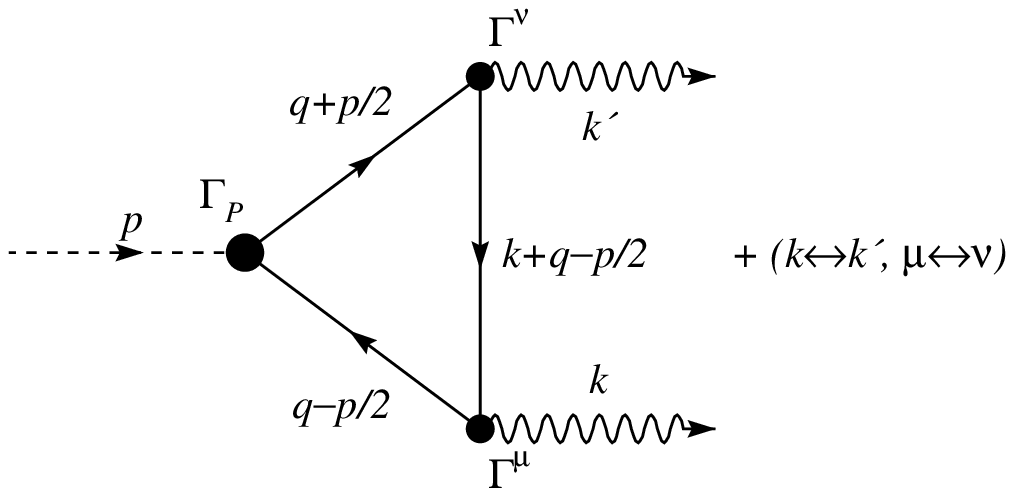}

\newpage

\vspace*{2cm}
\epsfxsize = 16 cm \epsfbox{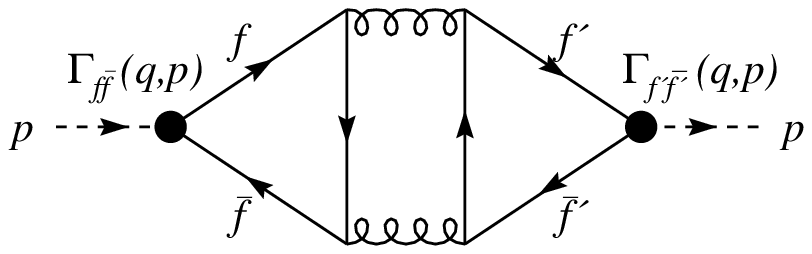}

\end{document}